\documentclass[journal]{IEEEtran}
\usepackage{times}
\usepackage{epsfig}
\usepackage{graphicx}
\usepackage{amsmath}
\usepackage{amssymb}
\usepackage{mathtools}
\usepackage{calligra}
\usepackage{color}
\usepackage{array}
\usepackage{hhline}
\usepackage{subfig}
\usepackage{multirow}
\usepackage{boxhandler}
\usepackage{tabularx}
\usepackage{adjustbox}
\usepackage{verbatim}
\usepackage{bm}
\usepackage{float}
\usepackage{amssymb}
\usepackage{pifont}
\newcommand{\cmark}{\ding{51}}%
\newcommand{\xmark}{\ding{55}}%

\usepackage{soul}
\soulregister\cite7 
\soulregister\citep7 
\soulregister\citet7 
\soulregister\ref7 
\soulregister\pageref7 

\begin{document}
	\title{Cross-SRN: Structure-Preserving Super-Resolution Network with Cross Convolution}
	\author{
		Yuqing~Liu,
		Qi~Jia,
		Xin Fan,~\IEEEmembership{Senior Member,~IEEE,}
		Shanshe Wang,
		Siwei~Ma,~\IEEEmembership{Member,~IEEE,}
		and~Wen~Gao,~\IEEEmembership{Fellow,~IEEE}
		\thanks{Y. Liu is with the School of Software, Dalian University of Technology,
			Dalian 116620, China (e-mail:liuyuqing@mail.dlut.edu.cn).}
		\thanks{Q. Jia and X. Fan are with International School of Information Science and Engineering, Dalian University of Technology, Dalian 116620, China (e-mail:
			jiaqi@dlut.edu.cn; xin.fan@dlut.edu.cn).}
		\thanks{S. Wang, S. Ma, and W. Gao are with the School of Electronics Engineering and Computer Science, Institute of Digital Media, Peking University, Beijing 100871, China (e-mail: sswang@pku.edu.cn; swma@pku.edu.cn).}
		\thanks{W. Gao is with the School of Electronics Engineering and Computer Science, Institute of Digital Media, Peking University, Beijing 100871, China, and Peng Cheng Laboratory, Shenzhen 518000, China (email: wgao@pku.edu.cn) }
		\thanks{Corresponding author: Qi Jia}
	}
	\markboth{Manuscript submitted to TCSVT}%
	{Shell \MakeLowercase{\textit{et al.}}: Bare Demo of IEEEtran.cls for IEEE Journals}
	\maketitle
	
\begin{abstract}
	It is challenging to restore low-resolution (LR) images to super-resolution (SR) images with correct and clear details. Existing deep learning works almost neglect the inherent structural information of images, which acts as an important role for visual perception of SR results. In this paper, we design a hierarchical feature exploitation network to probe and preserve structural information in a multi-scale feature fusion manner. First, we propose a cross convolution upon traditional edge detectors to localize and represent edge features. Then, cross convolution blocks (CCBs) are designed with feature normalization and channel attention to consider the inherent correlations of features. Finally, we leverage multi-scale feature fusion group (MFFG) to embed the cross convolution blocks and develop the relations of structural features in different scales hierarchically, invoking a lightweight structure-preserving network named as Cross-SRN. Experimental results demonstrate the Cross-SRN achieves competitive or superior restoration performances against the state-of-the-art methods with accurate and clear structural details. Moreover, we set a criterion to select images with rich structural textures. The proposed Cross-SRN outperforms the state-of-the-art methods on the selected benchmark, which demonstrates that our network has a significant advantage in preserving edges.
\end{abstract}


\IEEEpeerreviewmaketitle
	
\section{Introduction}
\IEEEPARstart{G}{iven} a low-resolution (LR) image, the task of super-resolution (SR) aims to find the corresponding high-resolution (HR) instance with refined details. Image SR has been considered in numerous computer vision tasks, such as recognition, person Re-Identification, semantic segmentation, and video compression.

As a highly ill-posed issue, image SR suffers from diverse degradation models, such as down-sampling, noise, and blur. Convolutional neural network (CNN) has shown its superior performance on complex information restoration~\cite{gao2021digital}, which is widely used by recent image SR works. SRCNN~\cite{srcnn_pami2016} is the first CNN-based image SR method with a three-layer network. Recently, some well-designed architectures, such as DRN~\cite{drn_cvpr2020}, HAN~\cite{han_eccv2020}, and LatticeNet~\cite{latticenet_eccv2020}, achieve the state-of-the-art performances by building deeper or wider networks to explore features more effectively. However, they almost neglect the inherent structural information of images, such as the outer contour of objects, lines, and curves, which is a vital factor to evaluate the restoration quality.

Gradient information in fixed direction is crucial to detect structural information. Traditional edge detectors, such as Prewitt and Sobel~\cite{sobel}, design filter-like templates to explore gradients maps in horizontal and vertical directions. Canny~\cite{canny_pani1986} obtains more accurate edge map by introducing Gaussian filter and double thresholds. These traditional methods employ templates with fix parameters, which can only detect edges with specified intensity. Meanwhile, traditional edge detectors are sensitive to image scale changes. Recently, CNN-based edge detectors achieve the state-of-the-art performances by learning filters on multi-scale images. BDCN~\cite{bdcn_pami2020} devices a bi-directional cascade network for perceptual edge detection. RCF~\cite{rcf_pami2019} considers the multi-scale hierarchical edge detection by fusing the features from different stages derived from VGG-16 backbone. 

According to the mechanism of human visual system (HVS), compared with other components, human eyes are most sensitive to the edge information on the image~\cite{r1q1_1}. As such, edge information is a vital important characteristic for vision~\cite{r1q1_2}. The visual quality of image is highly correlated with the edge information~\cite{r1q1_3}. As one of the semantic visual information, the reconstruction on edge maps represents the capacity of the method on detail restoration~\cite{r1q1_4}. A clear and accurate edge map shows that the high-quality restored image is with few artifacts~\cite{r1q1_5, r1q1_6}. There are also works concentrating on edge-preserving image super-resolution for better visualization performance~\cite{r1q1_5, r1q1_6, r1q1_7}.
SeaNet~\cite{seanet_tip2020} restores the HR images with a branch to recover the edges. DEGREE~\cite{degree_tip2017} introduces the loss between edges of LR and HR images for high-frequency information recovery. However, existing works only use edge maps as constraints while neglecting to construct specific filters or components to explore structure information directly.

\begin{figure}[t]
	\captionsetup[subfloat]{labelformat=empty, justification=centering}
	\begin{center}
		\newcommand{\rowArg}{1.65cm}
		\newcommand{\fullSize}{4.45cm}
		\newcommand{\fullWidth}{5.7cm}
		\newcommand{\patchSize}{1.85cm}
		\scriptsize
		\setlength\tabcolsep{0.05cm}
		\begin{tabular}[b]{c c c c}
			\multicolumn{3}{c}{\multirow{2}{*}[\rowArg]{
					\subfloat[Example image from Urban100~\cite{urban100}]
					{\includegraphics[height=\fullSize, width=\fullWidth]
						{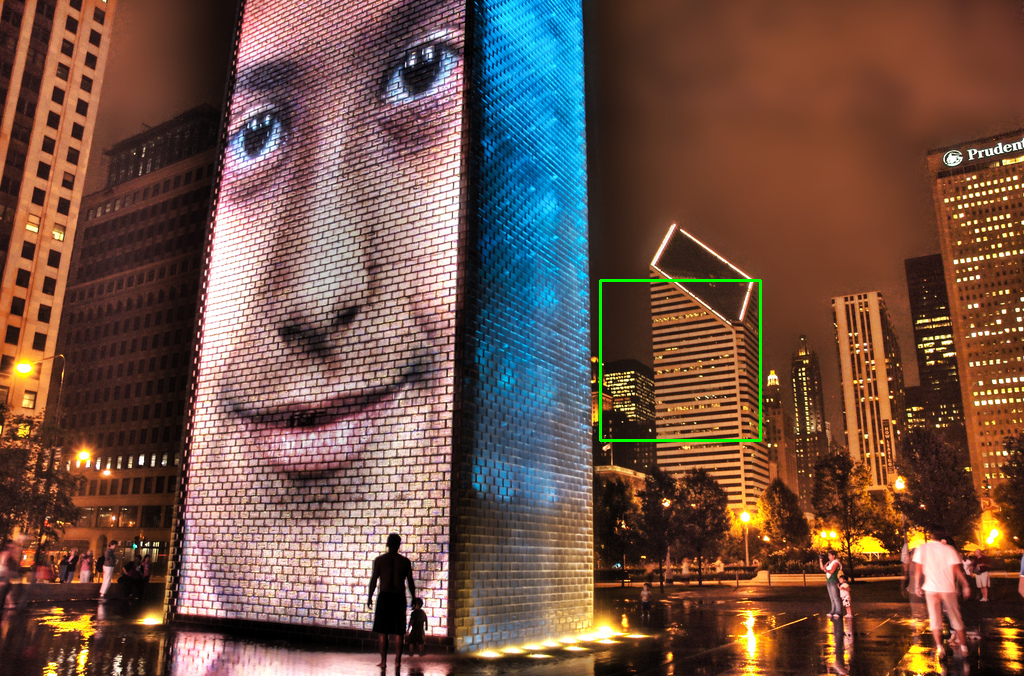}}}} &
			\subfloat[HR~\protect\linebreak(PSNR/SSIM)]
			{\includegraphics[width = \patchSize, height = \patchSize]
				{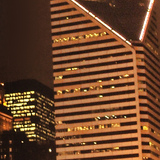}} \\ 			
			& & &
			\subfloat[LR~\protect\linebreak(20.31/0.7071)]
			{\includegraphics[width = \patchSize, height = \patchSize]
				{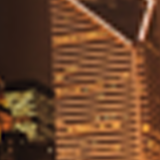}} \\ 	
			\subfloat[LapSRN~\cite{lapsrn_cvpr2017} \protect\linebreak(20.71/0.7574)]
			{\includegraphics[width = \patchSize, height = \patchSize]
				{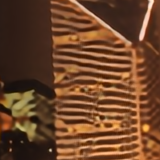}} &
			\subfloat[CARN~\cite{carn_eccv2018} \protect\linebreak(21.27/0.7785)]
			{\includegraphics[width = \patchSize, height = \patchSize]
				{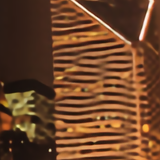}} &
			\subfloat[MSRN~\cite{msrn_eccv2018} \protect\linebreak(21.77/0.7990)]
			{\includegraphics[width = \patchSize, height = \patchSize]
				{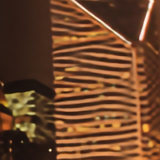}} &
			\subfloat[Cross-SRN\protect\linebreak(\textbf{21.99}/\textbf{0.8034})]
			{\includegraphics[width = \patchSize, height = \patchSize]
				{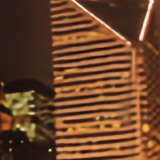}}
		\end{tabular}
	\end{center}
	\caption{Visual quality comparison for various image SR methods with scaling factor $\times4$.}
	\label{fig:slogan}
\end{figure}

Inspired by the edge detectors, this paper proposes a novel cross convolution to explore the structural information of features, which is composed of two factorized asymmetric filters. Two filters are applied simultaneously to increase the matrix rank and preserve more structural information. 
Based on the cross convolution, cross convolution block (CCB) is devised with feature normalization (F-Norm)~\cite{isrn} and CA~\cite{senet_pami2020} to consider the inherent correlations of features, which concentrate on the spatial and channel-wise information separately. 
CCBs are grouped in a multi-scale manner for feature exploration, termed as MFFG. MFFG considers the hierarchical edge information and progressively explore the structural information, where a padding structure is utilized to fully explore the features and residual connection is considered to keep the information. The MFFG modules are cascaded to constitute the final Cross-SRN. Experimental results show Cross-SRN achieves competitive or better performance than other works with more accurate structural information. Figure~\ref{fig:slogan} shows the visual quality comparison for various image SR methods, where Cross-SRN restores more correct line and edge textures. To demonstrate the restoration performance quantitatively, we build a selected benchmark from numerical existing textured HR images with plentiful structural information, which shows our network can preserve the structural textures effectively.

Our contributions can be concluded as follows:
\begin{itemize}
	\item Motivated by the edge detection methods, we design a cross convolution for effective structural information exploration.
	
	\item We devise a cross convolution block (CCB) to learn the relations of the edge features, and embedding the CCBs in a multi-scale feature fusion manner (MFFG) to explore the hierarchical features in different feature scales. 
	
	\item The proposed Cross-SRN achieves competitive or better performance against the stat-of-the-art methods with more accurate edge restoration. Especially, our network has a significant advantage over the selected benchmark with plentiful structure information.
\end{itemize}

\section{Related Works}
\subsection{Deep Learning for Image Super-Resolution}
CNN has proved to be an effective method for image restoration. SRCNN~\cite{srcnn_pami2016} is the first deep learning method for image SR, which contained three convolutional layers to present a sparse-coding like architecture. Then, VDSR~\cite{vdsr_cvpr2016} introduced a very deep network with residual learning for restoration. EDSR~\cite{edsr_cvpr2017} removed batch normalization and built a large network with residual blocks. Recently, CNN-based works mainly focus on building an effective mapping from LR to HR with elaborate blocks and architectures. Inspired by Laplacian pyramid, LapSRN~\cite{lapsrn_cvpr2017} and MS-LapSRN~\cite{lapsrn_pami2019} built the networks to restore the images with different scaling factors simultaneously. Ahn \textit{et al.} devised a cascading block in CARN~\cite{carn_eccv2018} for fast and accurate image restoration. MSRN~\cite{msrn_eccv2018}, RCAN~\cite{rcan_eccv2018}, RDN~\cite{rdn_pami2020}, and other recent works also achieved state-of-the-art performances with well-designed blocks. However, these works seldom address the structural information exploration.

There are works considering the edge and gradient map as a prior for restoration. DEGREE~\cite{degree_tip2017} introduced the gradient loss between HR and SR images to constrain the structural information generation. Ma \textit{et al.} investigated a GAN-based structure with gradient guidance in SPSR~\cite{spsr_cvpr2020}. Fang \textit{et al.} regarded edge as a prior in SeaNet~\cite{seanet_tip2020} to restore the high-frequency information. These works consider the edge or gradient as a guidance or prior to restore the structural information, but almost neglect to explore edge information directly.

Information distillation provides an efficient way for feature exploration, which is usually implemented in a multi-scale feature fusion manner. As far as we know, IDN~\cite{idn_cvpr2018} proposed by Hui~\textit{et al.} is the first SR network with information distillation, which utilized channel separation to distill the important features. Hui~\textit{et al.} proposed IMDN with multi-distillation and channel attention for better performances. RFDN~\cite{rfdn_eccv2020} was derived from IMDN and improved the network structure for fast and accurate restoration. However, these works seldom consider edge features in different scales upon limited computation costs.

\subsection{Attention Mechanism}
Attention mechanism is well established to emphasise vital information, which acts in a weighting distribution manner. Channel attention~\cite{senet_pami2020} (CA), as an efficient design for image SR, has been applied in recent state-of-the-art SR works. As far as we know, SENet~\cite{senet_pami2020} is the first work which introduces CA into deep learning. Recently, numerous works with CA has shown state-of-the-art performances on image SR. RCAN~\cite{rcan_eccv2018} investigated residual-in-residual blocks with CA to improve the performance. Dai~\textit{et al.} considered both CA and non-local attention in SAN~\cite{san_cvpr2019}. IMDN~\cite{imdn_mm2019}, RFDN~\cite{rfdn_eccv2020}, HAN~\cite{han_eccv2020} and DRN~\cite{drn_cvpr2020} also demonstrates superior performances with CA. However, existing CA estimates the information by global average pooling, neglecting the spatial relation between features.

\begin{figure*}[t]
	\centering
	\includegraphics[width=.8\linewidth]{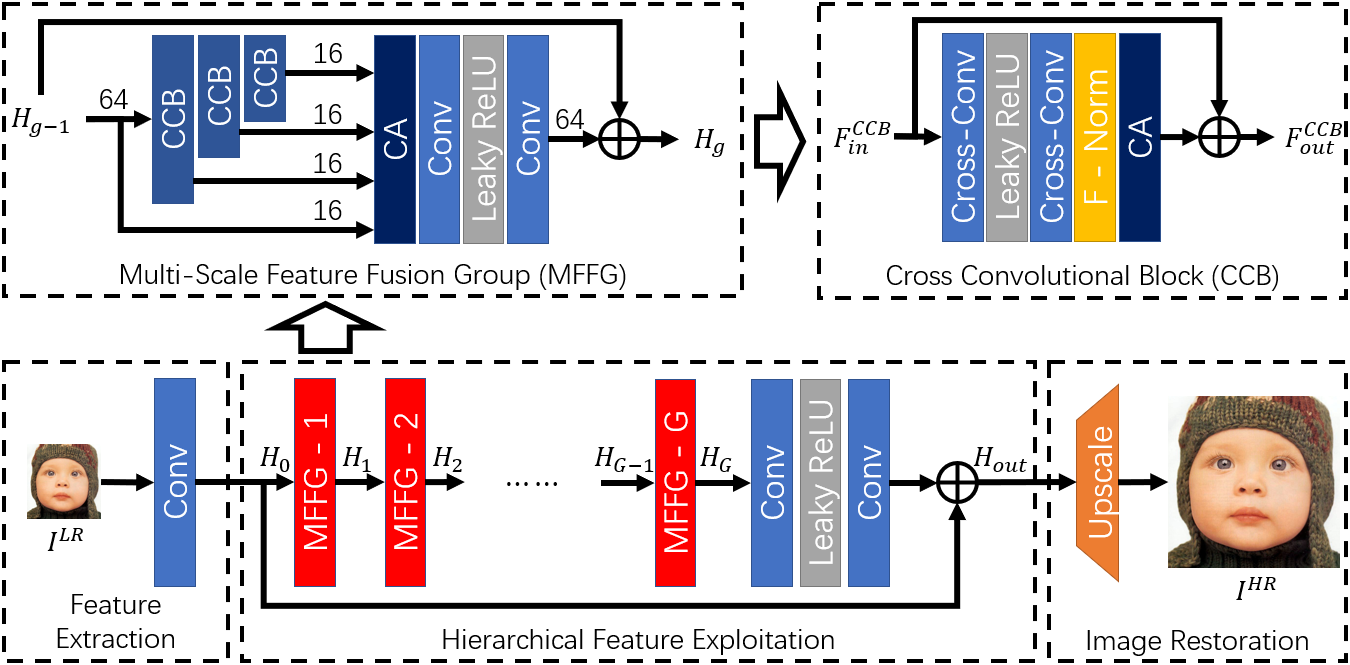}
	\caption{Network structure of Cross-SRN. The holistic network Cross-SRN is illustrated in the image below, including three steps: feature extraction, hierarchical feature exploitation, and image restoration. The detail of multi-scale feature fusion group (MFFG) in red and the emended cross convolutional block (CCB) in dark blue are demonstrated in the upper left and upper right, respectively.}
	\label{fig:network}
\end{figure*}

\section{Methodology}
In this section, we provide an overview of the proposed method. Then, we introduce the cross convolution and multi-scale feature fusion group (MFFG) modules in detail.

\subsection{Network Design}
An overview of the network design is shown in Figure~\ref{fig:network}. There are three steps in Cross-SRN, termed as feature extraction, hierarchical feature exploration, and image restoration, separated by different dashed boxes. Let $I^{LR}$ and $I^{HR}$ denote LR and HR instances, respectively. First, we use a convolutional layer to extract the original features from $I^{LR}$ as
\begin{equation}
	H_0 = f_{FE}(I^{LR}),
\end{equation}
where $f_{FE}(\cdot)$ denotes the feature extraction step. The convolution expands the channel number of the input instance and maps the instance into a specific space containing more potential information than the RGB space.

Then, we try to retain valuable information in different scales of features, while emphasizing the edge features. Thus, we design $G$ cascaded MFFGs for hierarchical feature exploration with global residual learning. There is,
\begin{equation}
	\begin{aligned}
		H_{g} & = f_{MFFG}^g(H_{g-1}),
	\end{aligned}
\end{equation}
where $f_{MFFG}^g(\cdot)$ denotes the $g$-th MFFG, and $H_{g-1}$ and $H_{g}$ denote the input and output features of the MFFG module, respectively. The final output of the cascaded MFFGs $H_G$ is fed into the residual module, which is composed of two convolutional layers with LeakyReLU layer. The padding structure with residual learning  is devised as,
\begin{equation}
	H_{out} = f_{PAD}(H_G) + H_0,
\end{equation}
where $f_{PAD}(\cdot)$ denotes the padding.

Finally, the HR images are restored by one convolution and a sup-pixel convolution, as,
\begin{equation}
	I^{HR} = f_{IR}(H_{out}).
\end{equation}
The final convolution decreases the channel number to restore the HR image and maps the features into RGB space.

\begin{figure}[t]
	\captionsetup[subfloat]{labelformat=empty, justification=centering}
	\begin{center}
		\subfloat[(a)]{\includegraphics[width = 0.25\linewidth]{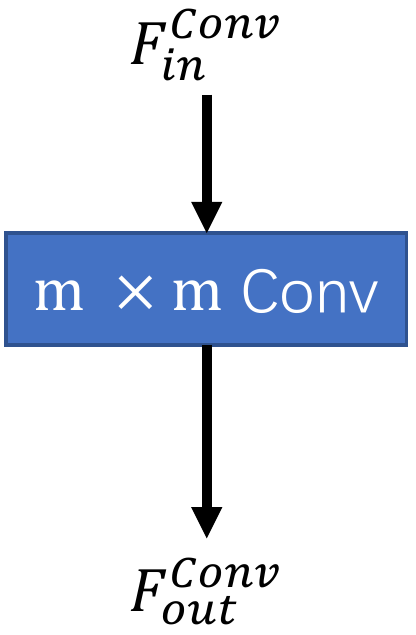}}\hspace{0.4cm}
		\subfloat[(b)]{\includegraphics[width = 0.55\linewidth]{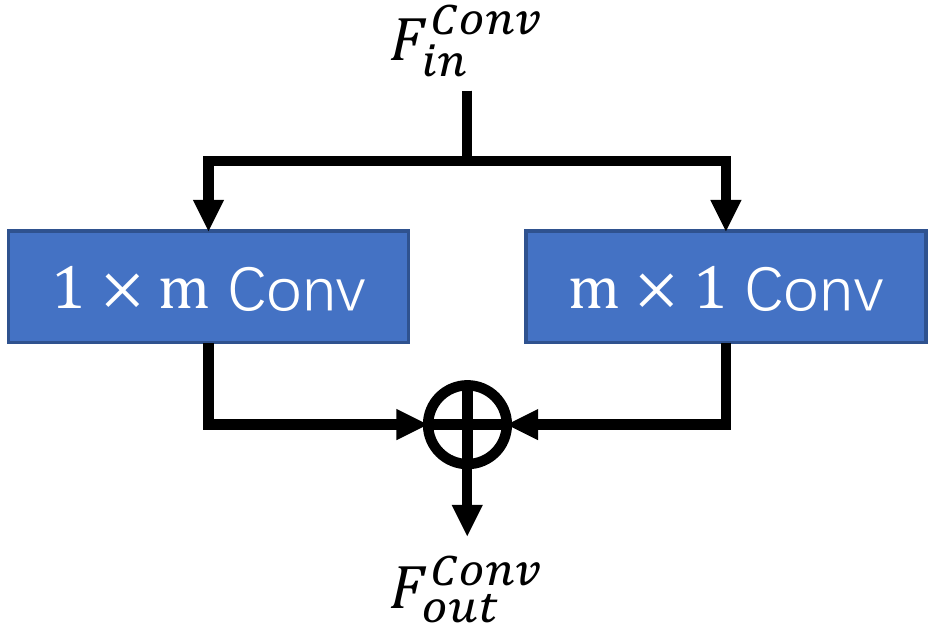}}
	\end{center}
	\vspace{-0.2cm}
	\caption{Illustration of proposed cross convolution. (a) Vanilla convolution. (b) Cross convolution. The input feature is parallelly processed by the factorized asymmetric filters, whose addition is regarded as the output.}
	\label{fig:crossconv}
\end{figure}

\subsection{Cross Convolution}
As shown in Figure~\ref{fig:crossconv}, different from vanilla convolutions, cross convolution is devised with two asymmetric perpendicular filters, which are denoted as $\mathbf{k}_{1 \times m}$ and $\mathbf{k}_{m \times 1}$ with receptive fields $1 \times m$ and $m \times 1$ separately. Let $F^{Conv}_{in}$, $F^{Conv}_{out}$ be the input and output feature, then there is,
\begin{equation}
	F^{Conv}_{out} = \mathbf{k}_{1 \times m} \otimes F_{in}^{Conv} + \mathbf{k}_{m \times 1} \otimes F_{in}^{Conv} + \mathbf{b},
\end{equation}
where $\otimes$ denotes the convolution, $\mathbf{b}$ is the bias term.

\begin{figure}[t]
	\centering
	\includegraphics[width=\linewidth]{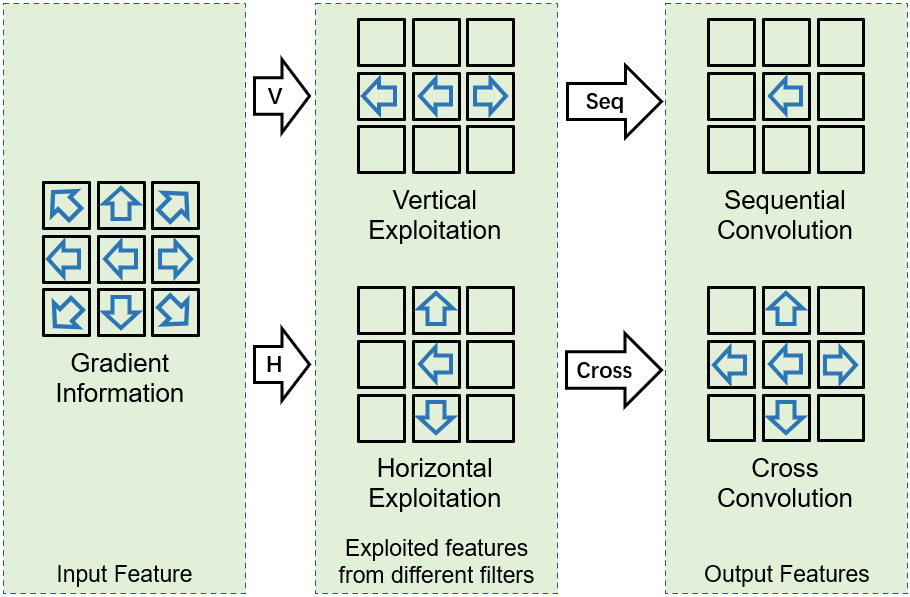}
	\caption{An example of the difference on information preservation between sequential and cross convolutions. The blue arrows demonstrate the gradient directions of every pixel. V and H denote the vertical and horizontal exploitation separately.}
	\label{fig:demo-cross}
\end{figure}

\begin{figure}[t]
	\captionsetup[subfloat]{labelformat=empty, justification=centering}
	\begin{center}
		\begin{tabular}[b]{c c c}
			\subfloat[(a) Origin]{\includegraphics[width = 0.28\linewidth]{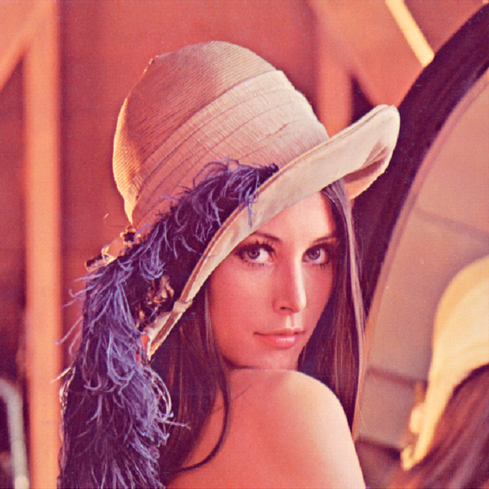}} &
			
			\subfloat[(b) Sobel]{\includegraphics[width = 0.28\linewidth]{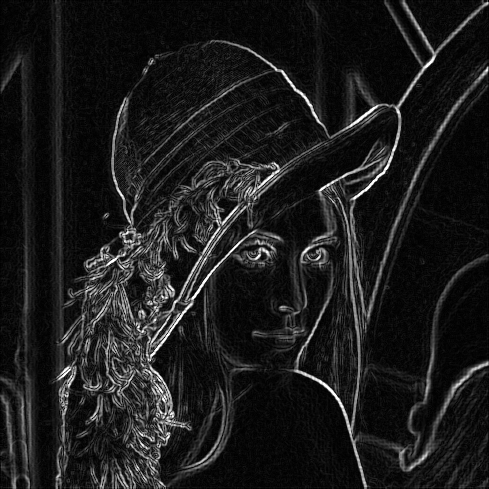}} &
			
			\subfloat[(c) Vertical]{\includegraphics[width = 0.28\linewidth]{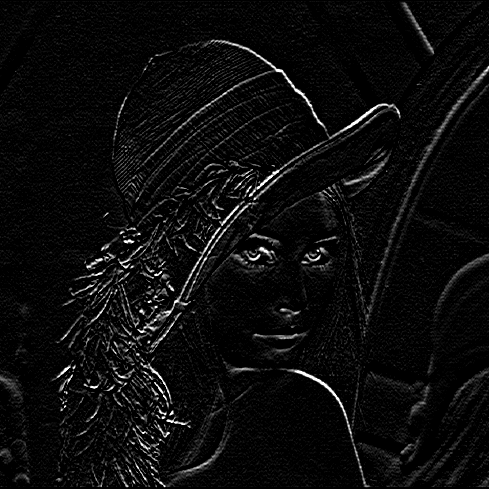}} \\
			
			\subfloat[(d) Horizontal]{\includegraphics[width = 0.28\linewidth]{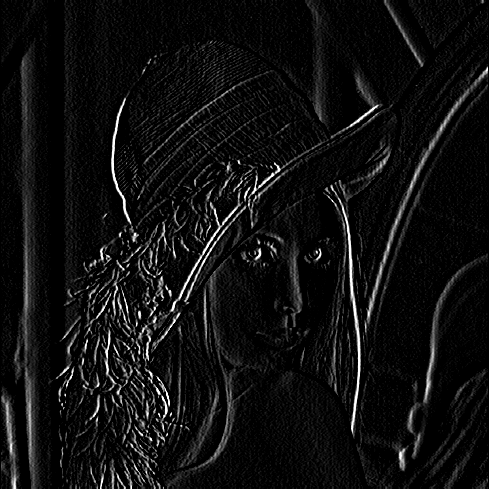}} &
			
			\subfloat[(e) Sequential]{\includegraphics[width = 0.28\linewidth]{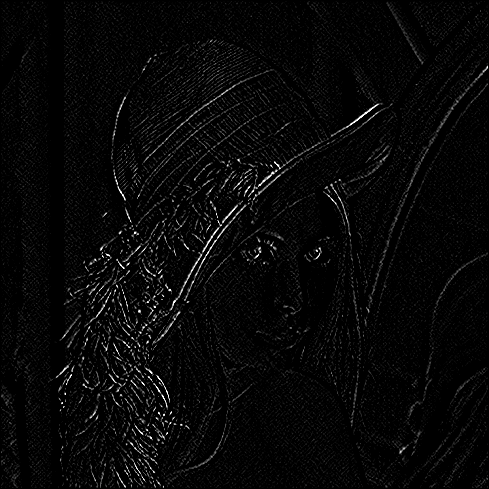}} &
			\subfloat[(f) Cross]{\includegraphics[width = 0.28\linewidth]{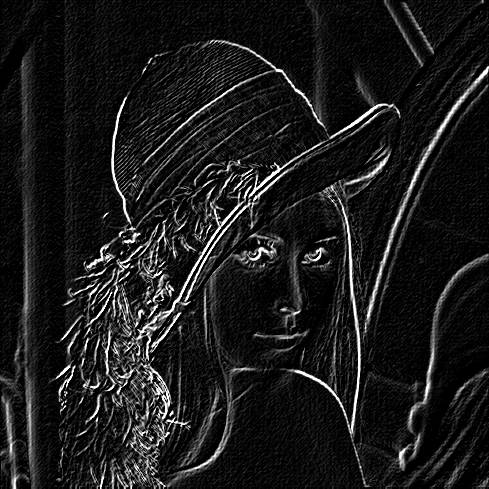}} \\
			
		\end{tabular}
	\end{center}
	\caption{An example of structural textures extracted by different filters. (a) Origin image. (b) Result of Sobel edge detector. (c) Edges extracted by vertical filter. (d) Edges extracted by horizontal filter. (e) Result of sequential convolution. (f) Result of cross convolution. }
	\label{fig:filter-crossconv}
\end{figure}

\begin{figure}[t]
	\centering
	\includegraphics[width=\linewidth]{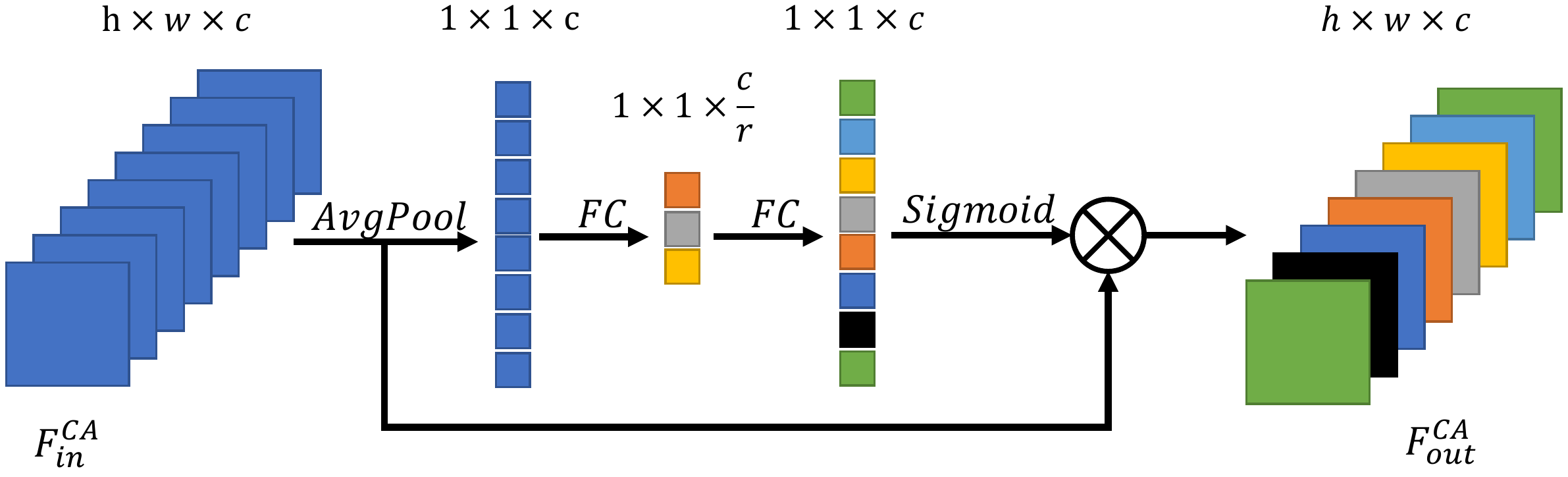}
	\caption{Illustration of channel attention~(CA).}
	\label{fig:ca}
\end{figure}

\begin{figure}[t]
	\centering
	\includegraphics[width=\linewidth]{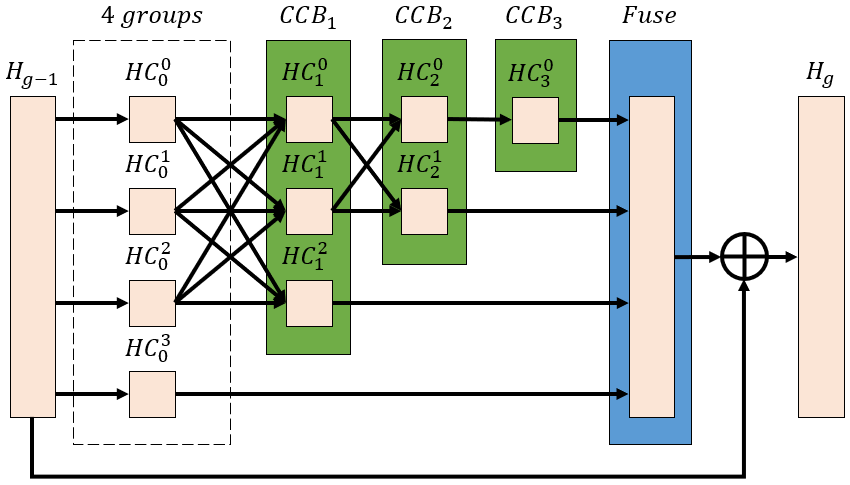}
	\caption{Illustration of multi-scale feature fusion group~(MFFG). The fuse structure is composed of two convolutional layers, one CA layer, and a Leaky ReLU.}
	\label{fig:idg-detail}
\end{figure}

Cross convolution emphasizes the edge information by exploiting the vertical and horizontal gradient information parallelly. The parallel design indicates fewer computation complexity and parameters than traditional filters with the same receptive field. Meanwhile, the parallel exploitation can also preserve more information than sequential design. Figure~\ref{fig:demo-cross} demonstrates the difference on information preservation between sequential and cross convolutions. The input feature contains gradient information in various directions, and the blue arrows demonstrate the gradient directions of every pixel. After vertical and horizontal exploration by two asymmetric filters, gradients from different directions are exploited. Finally, the output features with sequential convolution only focuses on the main gradient direction of the feature, while the cross convolution can preserve more gradient directions.

The advantage of the proposed cross convolution can be validated by the amount of information it holds. For a vanilla convolution filter $\mathbf{k}_{m \times m}$, the rank is $\mathbf{rank}(\mathbf{k}_{m \times m}) \leq m$. For the sequential combination of $\mathbf{k}_{1 \times m}$ and $\mathbf{k}_{m \times 1}$, the rank is $\mathbf{rank}(\mathbf{k}_{1 \times m} \cdot \mathbf{k}_{m \times 1})=1$. The filter with lower rank preserves less information than the filter with higher rank. The rank of cross convolution $\mathbf{k_{cross}}$ is $\mathbf{rank}(\mathbf{k}_{cross}) \leq 2$, which can preserve more latent information than the sequential one.


In fact, the cross convolution holds a similar formulation with the traditional edge detectors. Figure~\ref{fig:filter-crossconv} shows an example of structural textures extracted by different filters. We compare the cross convolution with Sobel operator, which detects the edges from both vertical and horizontal directions. Figure~\ref{fig:filter-crossconv} (b) and (f) are the results of Sobel and cross convolution, respectively. The comparison demonstrates that cross convolution can preserve most of the edge information with explicit and sharp edges, validating the capacity of structure texture exploration. Specially, we also compare the extracted edge maps with vertical, horizontal filters and sequential convolution, in figures (c), (d), and (e), which demonstrate that cross convolution can extract more edge information than other methods.

It is worth noting that Inception~\cite{inceptionV3} also has similar filters to Cross convolution, but their method is inapplicable to the challenging issues that we are addressing. Cross convolution is specially designed to explore edge information for image SR, while filters in Inception aim to save parameters for image classification, without explicit edge features. Cross convolution explores and overlays edge features in two perpendicular directions, invoking higher matrix rank of filters to preserve more information than channel aggregation of Inception. More importantly, Cross convolution is capable of any position of the networks for edge exploration, while Inception blocks can’t be used in the beginning of the network as stated in their paper. Furthermore, The sequentially stacked filters in Inception are discussed and termed as ``\textbf{Seq}'' in Figure~\ref{fig:demo-cross} and Section~\ref{sec:cross}. The contrast test in Table~\ref{tab:abl-crossconv} demonstrates Cross convolution preserves more information than \textbf{Seq}.

The cross convolution is designed with two factorized filters and concentrates on the vertical and horizontal structural information exploration. Similar to Canny, Sobel, and other traditional edge detectors, the factorized filters explore the edges in two orthogonal directions and derive the final structural information by summarization. As such, the cross convolution is inherently suitable for structural information exploration.

Separable convolution~\cite{separable} organizes the asymmetric filters in a sequential manner, while the proposed cross convolution designs the filters in a parallel fashion and achieves better quantitative performance with the same number of parameters. The motivation of spatial separable convolution is to convert the regular convolution into a parameter efficient design with less matrix multiplications and keep the receptive field. Compared with spatial separable convolution, the proposed cross convolution for image super-resolution only requires one extra addition operation with insignificant computational cost. The ``\textbf{Seq}'' in Table~\ref{tab:abl-crossconv} denotes the same design as the separable convolution. In the table, our cross convolution achieves better restoration performance with similar computational complexity. The cross convolution concentrates more on the structural information, which holds a similar formulation to the edge detectors with higher matrix rank and more potential information than the separable convolution.

\subsection{Multi-Scale Feature Fusion Group}

In order to obtain accurate edge information, we build the cross convolution block (CCB) based on the basic cross convolutions, as shown in Figure~\ref{fig:network}. Two cross convolutions with a Leaky ReLU activation are utilized to explore the structural information. Besides convolutions, F-Norm~\cite{isrn} and CA~\cite{senet_pami2020} are considered to emphasize important spatial and channel-wise information separately. Figure~\ref{fig:ca} shows the operation of CA, where global average pooling is applied to squeeze the information, and two full connection layers with ReLU activation explore the non-linear attention for every channel. F-Norm concentrates on the diversity of spatial information, which can be formulated as,
\begin{equation}
	F^{(i)}_{out} = (F^{(i)}_{in} \otimes \mathbf{k}^{(i)} + \mathbf{b}^{(i)} ) + F_{in}^{(i)},
\end{equation}
where $F^{(i)}_{in}$, $F^{(i)}_{out}$ are the input and output features of the $i$-th channel, $\mathbf{k}^{(i)}$ and $\mathbf{b}^{(i)}$ are the filter and bias for the $i$-th channel.

CCB in the network organizes the cross convolution in a residual block design. Residual block design has been widely considered in advanced networks for boosting the performance. Besides the residual connection, channel-wise attention and feature normalization are utilized to improve the exploration performance, which prove to be effective components for image SR.

As edge information is sensitive to scale changes, CCBs are grouped in a multi-scale feature fusion manner in MFFG to explore features in different scales. As shown in Figure~\ref{fig:idg-detail}, for the $g$-th MFFG, the input feature $H_{g-1}$ is divided into several groups with the same number of channels in average. As demonstrated in Figure~\ref{fig:idg-detail}, let $f_{CCB}^{j}(\cdot)$ be the $j$-th CCB in MFFG, and the input feature $H_{g-1}$ is divided into four groups from $HC_{0}^0$ to $HC_{0}^3$. The multi-scale feature fusion can be demonstrated as,
\begin{equation}
	\begin{aligned}
		&[HC_{1}^0, HC_{1}^1, HC_{1}^2] = f_{CCB}^1([HC_{0}^0, HC_{0}^1, HC_{0}^2]), \\
		&[HC_{2}^0, HC_{2}^1]           = f_{CCB}^2([HC_{1}^0, HC_{1}^1]), \\
		&[HC_{3}^0]                     = f_{CCB}^3([HC_{2}^0]), \\
	\end{aligned}
	\label{eq:idg}
\end{equation}
where $HC_{j}^k$ denotes the $k$-th group after the $j$-th CCB, and [$\cdot$] denotes the group combination.

The hierarchical features are aggregated by a residual block structure, which is composed of two convolutional layers with Leaky ReLU and one CA layer. Finally, the output of MFFG is,
\begin{equation}
	H_g = f_{Fuse}([HC_{3}^0, HC_{2}^1, HC_{1}^2, HC_{0}^3]) + H_{g-1},
\end{equation}
where $f_{Fuse}(\cdot)$ is the fusion structure.

Herein, MFFG keeps the original information and emphasizes the structural information hierarchically. As shown in the upper left of the Figure~\ref{fig:network}, the input feature $H_{g-1}$ are separated into four groups. Three groups are sequentially processed by CCBs for hierarchical structural information exploitation. In order to preserve the potential information lost during edges exploring, the last group keeps the identical original features. After the CCBs, a vanilla residual block structure with channel attention is utilized for effective feature exploration and gradient transmission.

\begin{figure}[t]
	\captionsetup[subfloat]{labelformat=empty, justification=centering}
	\begin{center}
		\subfloat[(a)]{\includegraphics[width = 0.5\linewidth]{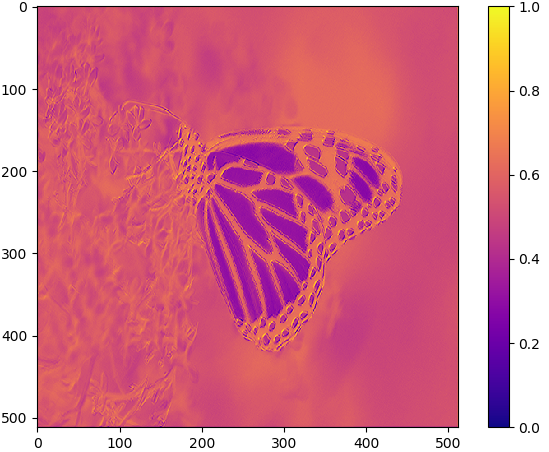}}
		\subfloat[(b)]{\includegraphics[width = 0.5\linewidth]{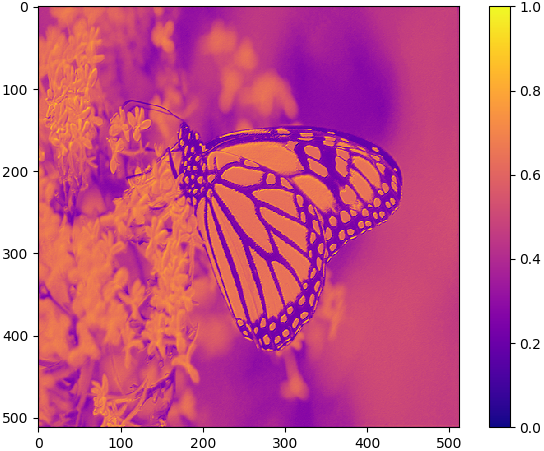}} \\
		\subfloat[(c)]{\includegraphics[width = 0.5\linewidth]{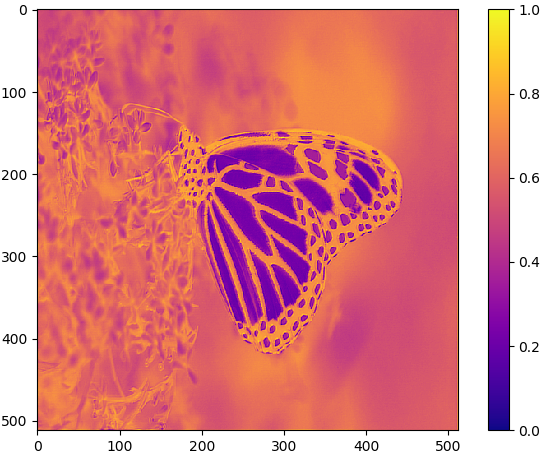}}
		\subfloat[(d)]{\includegraphics[width = 0.5\linewidth]{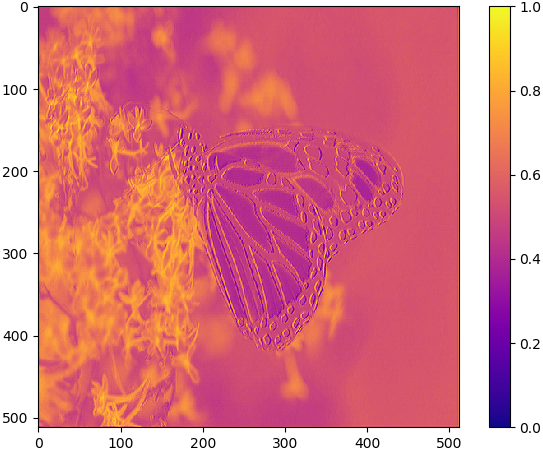}}
	\end{center}
	\caption{Visualized feature maps processed by different convolution designs. (a) Input feature. (b) Feature processed by the vertical filter. (c) Feature processed by the horizontal filter. (d) Output feature of cross convolution. The values are calculated by averaging the feature maps and normalized in range 0 to 1.}
	\label{fig:abl-crossconv}
\end{figure}

\begin{table*}[t]
	\centering
	\caption{Comparison of three different convolution designs on PSNR/SSIM with \textbf{BI} $\times4$ degradation.}
	\label{tab:abl-crossconv}
	\begin{tabular}{|c|c|c|c|c|c|c|c|c|}
		\hline
		\textbf{Method}& \textbf{Param}& \textbf{MACs}& \textbf{Selective}& \textbf{Set5}& \textbf{Set14}& \textbf{B100}& \textbf{Urban100}& \textbf{Manga109} \\
		\hline
		\textbf{Seq}  & 1,296K& 74.2G& 24.39/0.8527& 32.20/0.8949& 28.61/0.7817& 27.57/0.7361& 26.15/0.7875& 30.49/0.9077\\
		\textbf{Conv} & 1,509K& 86.5G& 24.44/0.8549& 32.20/0.8946& 28.60/0.7823& 27.58/0.7367& 26.17/0.7889& 30.54/0.9092\\
		\textbf{Cross}& {1,296K}& {74.2G}& 24.46/0.8539& {32.24/0.8954}& {28.59/0.7817}& {27.58/0.7364}& {26.16/0.7881}& {30.53/0.9081}\\
		\hline
	\end{tabular}
\end{table*}

\begin{table}[t]
	\centering
	\caption{Investigation on multi-scale feature fusion on PSNR/SSIM with \textbf{BI} $\times4$ degradation.}
	\label{tab:abl-idg}
	\fontsize{6.5}{8}\selectfont
	\begin{tabular}{|c|c|c|c|c|c|}
		\hline
		\textbf{Method}& \textbf{Param}& \textbf{MACs}& \textbf{Set5}& \textbf{Set14}& \textbf{B100}\\
		\hline
		\textbf{w/o MFF}   & 2,366K& 135.2G& 32.24/0.8957& 28.65/0.7837& 27.62/0.7379\\
		\textbf{w/o CCB}& 847K  & 48.5G & 32.04/0.8926& 28.48/0.7792& 27.50/0.7338\\
		\textbf{Ours}  & {1,296K}& {74.2G} & {32.24/0.8954}& {28.59/0.7817}& {27.58/0.7364}\\
		\hline
	\end{tabular}
\end{table}

\begin{table}[t]
	\centering
	\caption{PSNR/SSIM, parameters and MACs comparisons between sequential convolution and cross convolution on Set5 with scaling factor $\times4$.}
	\label{tab:abl-mxm}
	\fontsize{6.5}{8}\selectfont
	\begin{tabular}{|c|c|c|c|c|c|c|}
		\hline
		\textbf{Receptive Field}& \multicolumn{2}{c|}{$3\times3$}& \multicolumn{2}{c|}{$5\times5$}& \multicolumn{2}{c|}{$7\times7$} \\
		\hline
		\textbf{Convolution Type}& \textbf{Seq}& \textbf{Cross}& \textbf{Seq}& \textbf{Cross}& \textbf{Seq}& \textbf{Cross}\\
		\hline
		\textbf{PSNR}& 31.99& 32.07& 32.10& 32.14& 32.08& 32.08 \\
		\textbf{SSIM}& 0.8924& 0.8926& 0.8931& 0.8936& 0.8932& 0.8931 \\
		\textbf{Params (M)}& 1.29& 1.29& 1.58& 1.58& 1.87& 1.87 \\
		\textbf{MACs (G)}& 74.22& 74.22& 90.74& 90.74& 107.25& 107.25 \\
		\hline		
	\end{tabular}
\end{table}

\begin{table}[t]
	\centering
	\caption{Investigation on channel attention~(CA) and feature normalization~(F-Norm).}
	\label{tab:abl-ca}
	\begin{tabular}{|c|c|c|c|c|}
		\hline
		\textbf{CA}& \textbf{FN}& \textbf{Set5}& \textbf{Urban100}& \textbf{Manga109} \\
		\hline
		\cmark& \xmark& 32.20/0.8952& 26.15/0.7875& 30.47/0.9079\\
		\xmark& \cmark& 32.13/0.8948& 26.11/0.7872& 30.50/0.9086\\
		\cmark& \cmark& {32.24/0.8954}& {26.16/0.7881}& {30.53/0.9081}\\
		\hline
	\end{tabular}
\end{table}

\begin{figure*}[t]
	\captionsetup[subfloat]{labelformat=empty, justification=centering}
	\begin{center}
		\subfloat[(a)]{\includegraphics[width = 0.165\linewidth]{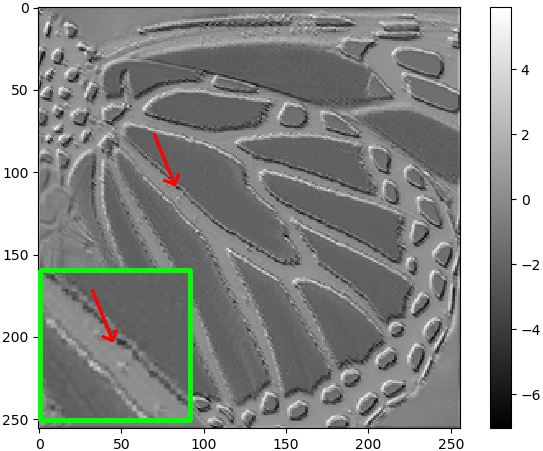}}
		\subfloat[(b)]{\includegraphics[width = 0.165\linewidth]{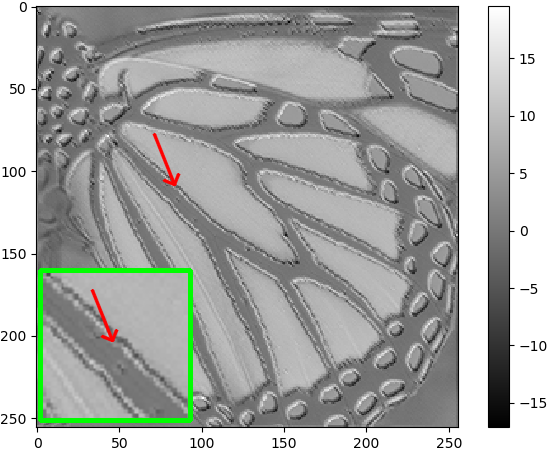}}
		\subfloat[(c)]{\includegraphics[width = 0.165\linewidth]{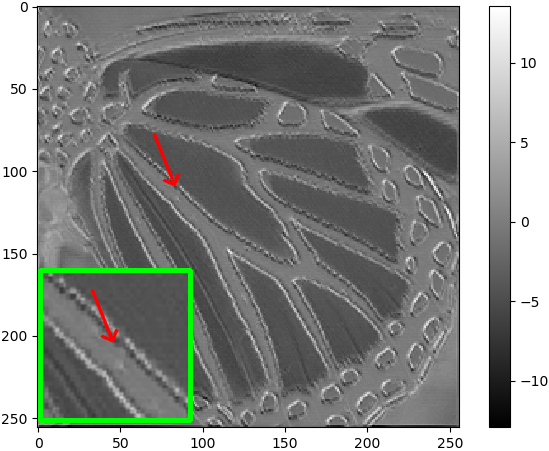}}
		\subfloat[(d)]{\includegraphics[width = 0.165\linewidth]{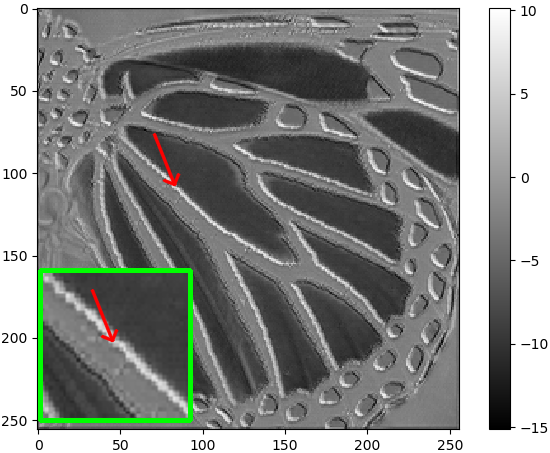}}
		\subfloat[(e)]{\includegraphics[width = 0.165\linewidth]{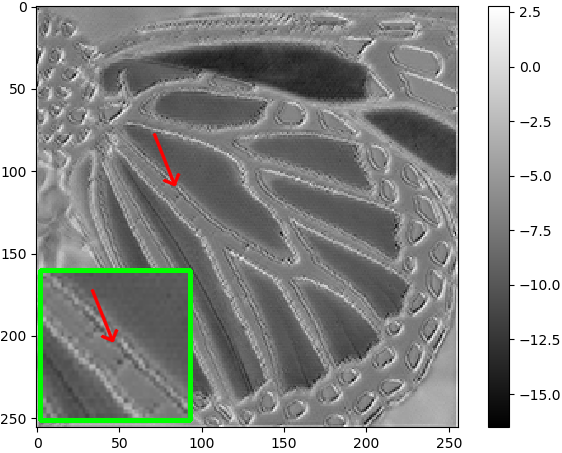}}
		\subfloat[(f)]{\includegraphics[width = 0.165\linewidth]{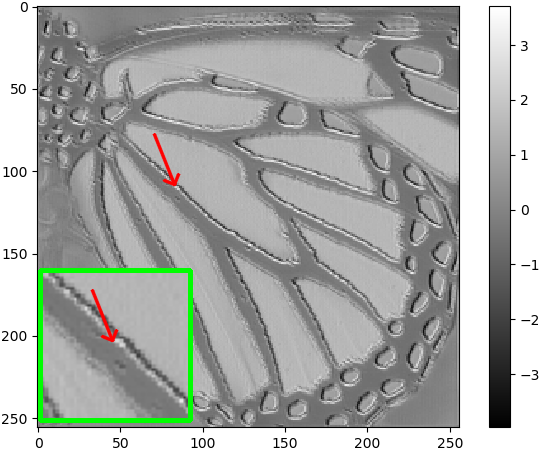}}
	\end{center}
	\caption{Visualized feature maps of multi-scale feature fusion. (a): input feature. (b)-(e): different groups after CCBs, which denote from $HC_0^3$ to $HC_3^0$ in Eq.~\ref{eq:idg}. (f): output feature. All the images are illustrated by the same gray-scale strategy.}
	\label{fig:abl-distillation}
\end{figure*}

\section{Experiments and Analysis}
In this section, we introduce the experiment settings of our Cross-SRN in the beginning. Then, we provide the ablation study on the cross convolution and MFFG to show the effectiveness of the proposed model. Finally, we compare our Cross-SRN with the state-of-the-art works.

\subsection{Experiment Settings}
In Cross-SRN, there are $G=10$ MFFGs for non-linear exploitation. Filter numbers of convolution layers are set as $c=64$ except for the image restoration module, and the kernel sizes of filters are set as $m=3$. Since the largest scaling factor in this paper is set as $\times4$, the image restoration module can be regarded as a down-sampling step from the feature space to RGB space. We train the network with DIV2K~\cite{div2k} dataset. DIV2K is firstly proposed in New Trends in Image Restoration and Enhancement (NTIRE) 2017 competition and has been widely used for image SR tasks. We choose 895 images for training, and 5 images for validation. We leverage bicubic~\cite{bicubic} (\textbf{BI}) to obtain the degraded images by different scaling factors. For training sets, images are cropped with patch size $48\times48$, and then randomly flipped and rotated for augmentation. Cross-SRN is updated by an Adam~\cite{adam_iclr2015} optimizer with learning rate $lr=10^{-4}$. We train the network for 1000 epochs, and halve the learning rate for every 200 epochs. The evaluation indicators are chosen as peak signal-to-noise ratio (PSNR), structural similarity (SSIM) and multiply-accumulate operations (MACs). The higher PSNR/SSIM result means better performance, and the lower MACs means faster speed. MACs is calculated by restoring a $1280\times720$ (720P) image with scaling factor $\times4$. 

Five testing benchmarks are used for performance comparison: Set5~\cite{set5}, Set14~\cite{set14}, B100~\cite{b100}, Urban100~\cite{urban100}, and Manga109~\cite{manga109}. Urban100 is a testing benchmark with real-world HR images. B100 is also a benchmark from real-world with abundant complex textures. Manga109 is composed of comic covers. All of the three benchmarks contain plentiful high-frequency and structural information. Furthermore, we build a selective benchmark to compare the capacities of structural information recovery, which is composed of images from Urban100 and Manga109 with numerical edges and lines. The benchmark is built according to the following steps. Firstly, we blur the images with $7\times7$ Gaussian kernel for denoising. After filtering, Sobel operator is utilized to extract the edges from the images. We use threshold $t_e = 128$ to remove the weak responses from edge maps, and calculate the average response of image. The images with responses higher than $t_r=12$ are included in the benchmark. There are 38 images in total, including 18 images from Urban100 and 20 images from Manga109.

$t_e$ and $t_r$ are used to select images with more edge information. $t_e$ is used to explore edges with high response and $t_r$ is used to evaluate the average intensity of edge information in images. The edge map extracted by Sobel represents the responding of edge information in every pixel with range 0-255. $t_e$ is set as 128, an average value of the range. Pixels with a value less than $t_e$ is set as 0. Then, the average response of each pixel is calculated to estimate the intensity of edge information in each image. If the response is larger than an empirical value $t_r$=12, the image is included in the selective benchmark.

\subsection{Ablation Study}
Herein, we investigate the performance gained from the cross convolution, the multi-scale feature fusion and the network structure separately.

\subsubsection{Investigation on Cross Convolution}
\label{sec:cross}
To demonstrate the performance of cross convolution, we design three convolution structures for comparison: \textbf{Cross} denotes the proposed cross convolution, \textbf{Seq} denotes the sequential convolution, and \textbf{Conv} denotes the vanilla convolution. To investigate the effectiveness of \textbf{Cross} on structural information preservation, we test the three convolution structures on the selective benchmarks and other widely used benchmarks. Table~\ref{tab:abl-crossconv} demonstrates the comparison on PSNR/SSIM. We can see \textbf{Cross} achieves competitive or better PSNR/SSIM performance on the testing benchmarks with fewer parameters and MACs, which proves our hypothesis on matrix rank. Specially, on the selective benchmark with plentiful edge structures, \textbf{Cross} achieves higher PSNR than \textbf{Conv} and \textbf{Seq}, which means the proposed \textbf{Cross} achieves better performance on structural information preservation with only 76.8\% MACs and 85.8\% parameters of \textbf{Conv} or \textbf{Seq}. Thus, \textbf{Cross} makes a good balance on edge feature exploration and information loss.

The visualized analysis of three convolution structures is demonstrated in Figure~\ref{fig:abl-crossconv}. In Figure~\ref{fig:abl-crossconv}~(b) and (c), the features processed by two factorized filters response to different kinds of areas. Figures (b) and (c) are processed by different directions of filters, which have high responses to two directional margins, respectively. On the contrary, the proposed cross convolution extracts edge features parallelly from the same input. As shown in Figure~\ref{fig:abl-crossconv}~(d), the edges are enhanced after cross convolution, which also demonstrates our capacity on structural texture restoration.

The cross convolution is designed to substitute the ordinary convolution for addressing the structural information with restricted parameters and MACs. A vanilla $m \times m$ convolutional filter can be regarded as a matrix with shape $m \times m$, and the rank of the filter $\mathbf{rank}(\mathbf{k}_{m \times m}) \leq m$. In other words, the upper bound of the filter’s rank is m. The cross convolution is designed as the combination of $\mathbf{k}_{m \times 1}$ and $\mathbf{k}_{1 \times m}$, whose upper bound is 2. As such, the cross convolution can preserve the most information of ordinary convolution when m is small. Furthermore, the cross convolution follows a similar formulation to the traditional edge detectors, which can effectively explore the structural information. The experimental results show cross convolution achieves competitive PSNR/SSIM performance than the ordinary convolution with fewer parameters and MACs.

The sequential situation \textbf{Seq} holds a same receptive field as the vanilla convolution, but the sequential exploration losses much information. The rank of sequential convolution $\mathbf{rank}(\mathbf{k}_{1 \times m} \cdot \mathbf{k}_{m \times 1})=1$, which means the sequential convolution potentially losses half of the information than the cross convolution.

To better illustrate the potential information loss, we perform the ablation studies to compare the PSNR/SSIM. The results are shown in Table.~\ref{tab:abl-mxm}

Table~\ref{tab:abl-mxm} shows the PSNR/SSIM, parameters and MACs comparisons between sequential convolution and cross convolution on Set5 benchmark with scaling factor $\times4$. For a fair comparison, all models are re-trained for 200 epochs. In the table, we can find that the cross convolution achieves better PSNR/SSIM performance than the sequential design on all receptive fields. When the receptive field is $3\times3$, there is 0.8 dB PSNR improvement, which is significant. This observation is in accordance with our conclusion on the rank of filters which shows the cross convolution can preserve more information than the sequential design. Furthermore, we can find that the performances of cross convolution are similar when the receptive fields are different. As we discussed on the filter’s rank, the upper bound of cross convolution’s rank is 2. So, choosing filter size $m=3$ is the most efficient way for restoring the HR image.

There is another reason for choosing $m=3$. According to the linearity of convolution operation, the filter with larger receptive fields can be equivalently substituted by a combination of filters with smaller receptive fields~\cite{inceptionV3}. The receptive filed $3\times3$ is the economic choice to build a deeper neural network, which has been widely considered in recent SR works~\cite{edsr_cvpr2017, rdn_pami2020, rcan_eccv2018}.

\subsubsection{Investigation on Multi-Scale Feature Fusion} To investigate the effectiveness of the multi-scale feature fusion module, we compare the proposed MFFG design with the other two optional designs, one is MFFG without channel division and multi-scale feature fusion, which is denoted as~\textbf{w/o MFF}. The other is MFFG without CCBs, which is termed as~\textbf{w/o CCB}. 
The \textbf{w/o CCB} means we remove all the CCBs from the Cross-SRN. In other words, the three CCBs in MFFG are omitted while other components are not modified. The ablation study of \textbf{w/o CCB} aims to investigate the effectiveness of the backbone of Cross-SRN.
In the~\textbf{w/o MFF} version, the three CCBs all take all the feature channels as input, and all the CCBs are processed sequentially. 
In other words, \textbf{w/o MFF} means all the CCBs in MFFG are modified with channel number as 64, and no channel separation is considered. The MFFG without multi-scale feature fusion can be regarded as a residual-in-residual-like architecture by stacking three CCBs, two convolutional layers, one LeakyReLU and one CA.
The comparison results on PSNR/SSIM are illustrated in Table~\ref{tab:abl-idg}. We can see \textbf{Ours} achieves competitive performance than \textbf{w/o MFF} with around half parameters and MACs. \textbf{Ours} drops less than 0.01 dB on Set5 and 0.05db on B100. The model without MFF holds similar PSNR/SSIM to MFFG but the Params and MACs are much higher. The results demonstrate that multi-scale feature fusion preserves most edge information by hierarchical exploration. The model without CCB drops near 0.2 dB on Set5 when compared with MFFG, which means CCB is crucial for restoration.

\begin{figure*}[t]
	\captionsetup[subfloat]{labelformat=empty, justification=centering}
	\begin{center}
		\subfloat[HR~(PSNR/SSIM)]{\includegraphics[width = 0.2\linewidth]{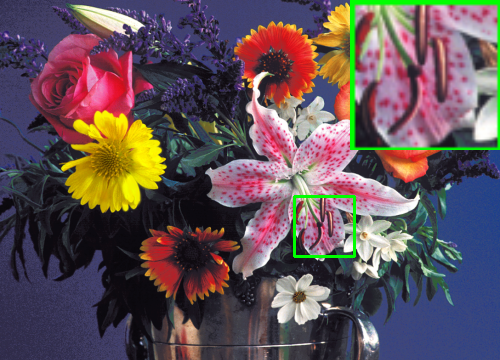}}
		\subfloat[LR~(23.67/0.6932)]{\includegraphics[width = 0.2\linewidth]{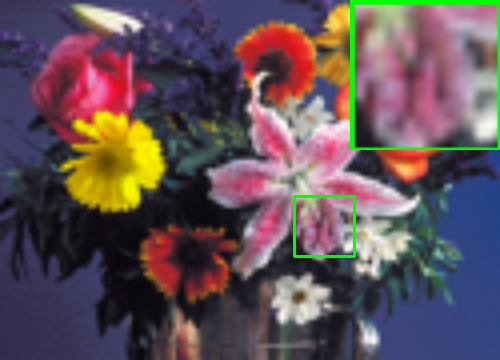}}
		\subfloat[MSLapSRN~(26.71/0.8119)]{\includegraphics[width = 0.2\linewidth]{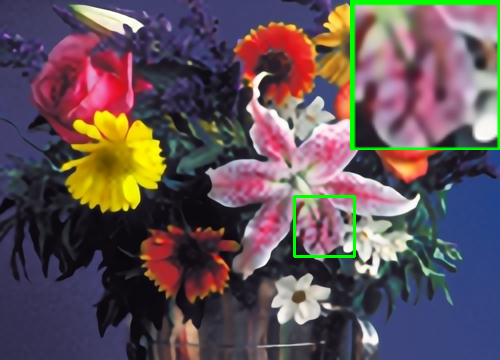}}
		\subfloat[IMDN~(27.03/0.8212)]{\includegraphics[width = 0.2\linewidth]{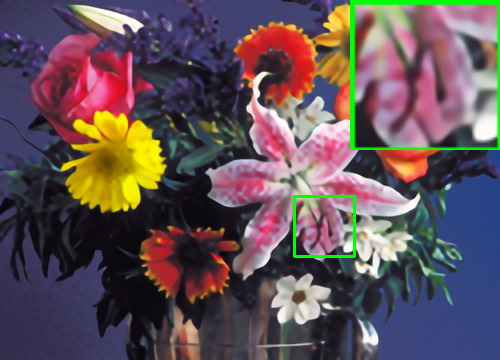}}
		\subfloat[Ours~(\textbf{27.04/0.8220})]{\includegraphics[width = 0.2\linewidth]{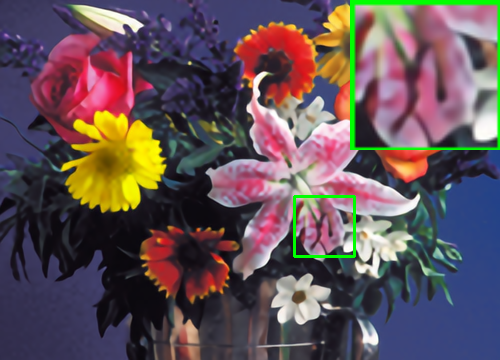}} \\
		
		\subfloat[HR~(PSNR/SSIM)]{\includegraphics[width = 0.2\linewidth]{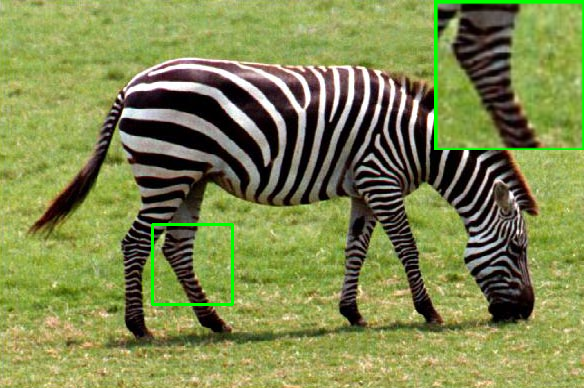}}
		\subfloat[LR~(21.90/0.6331)]{\includegraphics[width = 0.2\linewidth]{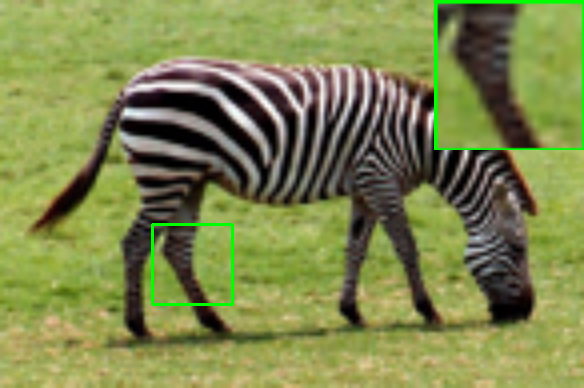}}
		\subfloat[MSLapSRN~(25.69/0.7771)]{\includegraphics[width = 0.2\linewidth]{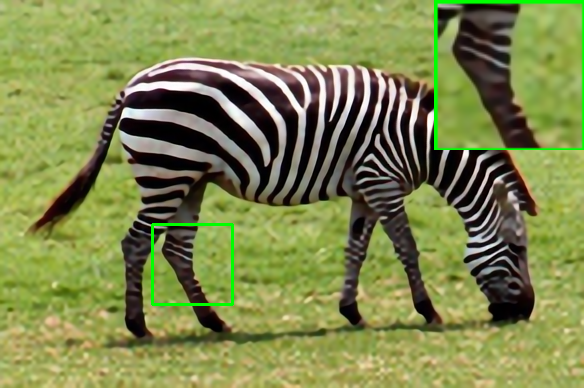}}
		\subfloat[IMDN~(26.29/0.7864)]{\includegraphics[width = 0.2\linewidth]{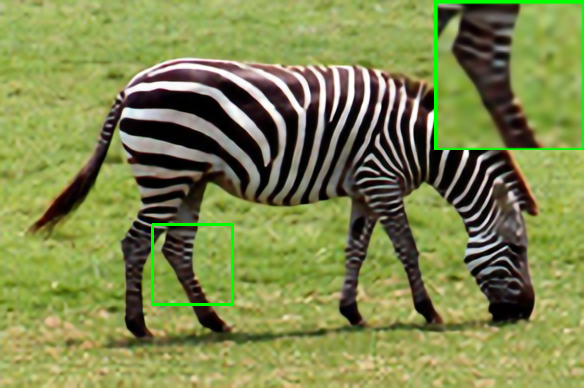}}
		\subfloat[Ours~(\textbf{26.31/0.7849})]{\includegraphics[width = 0.2\linewidth]{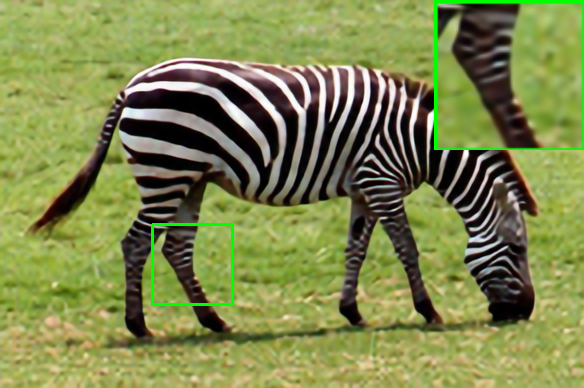}} \\
	\end{center}
	\vspace{-0.2cm}
	\caption{Visualization comparison on Set14 dataset.}
	\label{fig:set14}
\end{figure*}

\begin{figure*}[t]
	\captionsetup[subfloat]{labelformat=empty, justification=centering}
	\begin{center}
		\subfloat[HR~(PSNR/SSIM)]{\includegraphics[width = 0.16\linewidth]{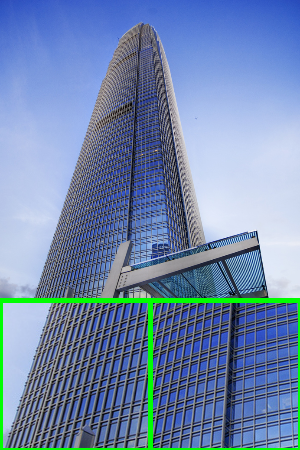}}
		\subfloat[LR~(21.47/0.7405)]{\includegraphics[width = 0.16\linewidth]{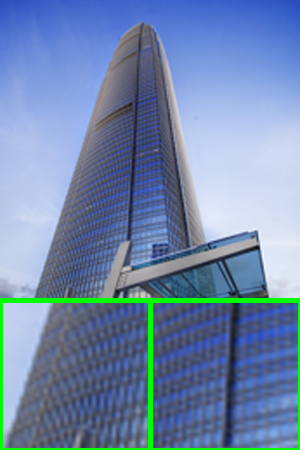}}
		\subfloat[VDSR~(22.04/0.7839)]{\includegraphics[width = 0.16\linewidth]{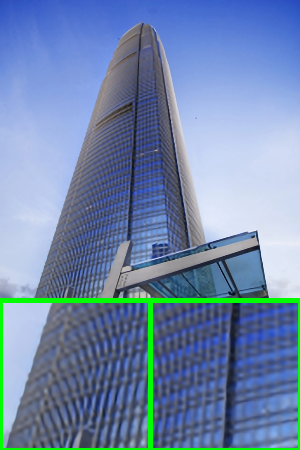}}
		\subfloat[CARN~(22.30/0.8003)]{\includegraphics[width = 0.16\linewidth]{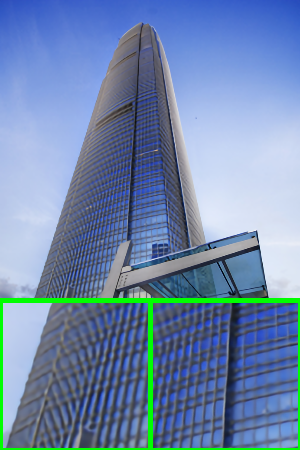}}
		\subfloat[IMDN~(22.30/0.8005)]{\includegraphics[width = 0.16\linewidth]{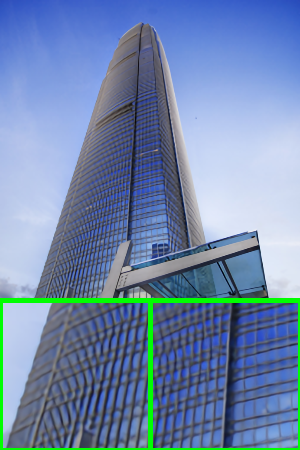}}
		\subfloat[Ours~(\textbf{22.43/0.8060})]{\includegraphics[width=0.16\linewidth]{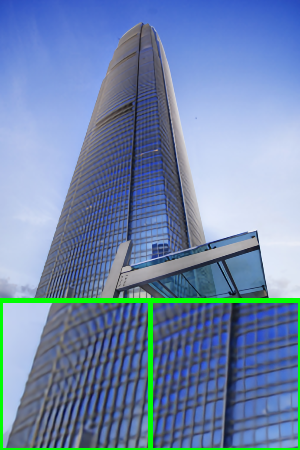}} \\
		\subfloat[HR~(PSNR/SSIM)]{\includegraphics[width = 0.16\linewidth]{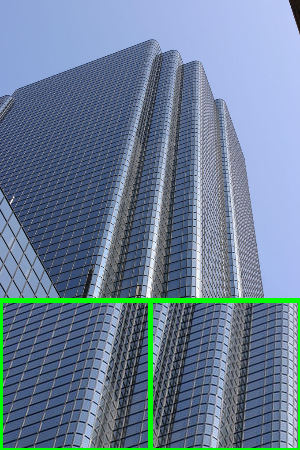}}
		\subfloat[LR~(20.91/0.6480)]{\includegraphics[width = 0.16\linewidth]{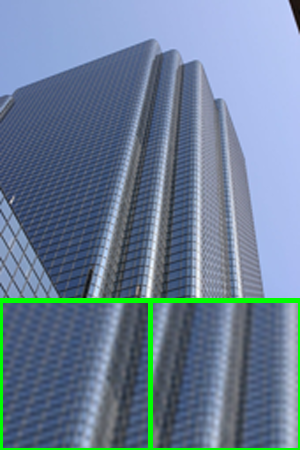}}
		\subfloat[VDSR~(21.81/0.7182)]{\includegraphics[width = 0.16\linewidth]{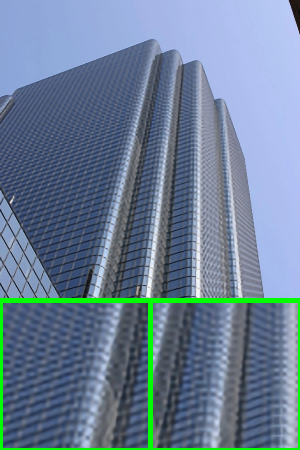}}
		\subfloat[CARN~(22.32/0.7612)]{\includegraphics[width = 0.16\linewidth]{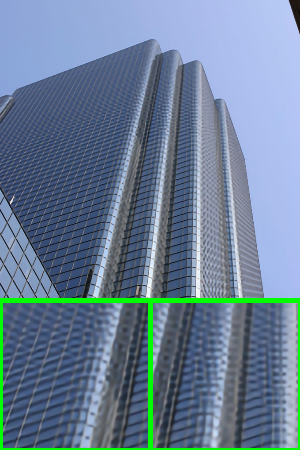}}
		\subfloat[IMDN~(22.48/0.7705)]{\includegraphics[width = 0.16\linewidth]{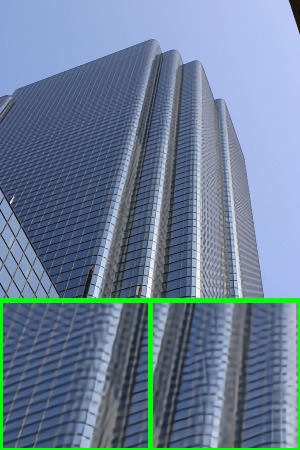}}
		\subfloat[Ours~(\textbf{22.50/0.7791})]{\includegraphics[width=0.16\linewidth]{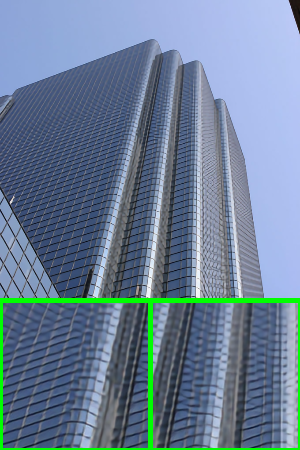}}
	\end{center}
	\vspace{-0.2cm}
	\caption{Visualization comparison on Urban100 dataset.}
	\label{fig:urban100}
\end{figure*}

\begin{figure*}[t]
	\captionsetup[subfloat]{labelformat=empty, justification=centering}
	\begin{center}
		\subfloat[HR~(PSNR/SSIM)]{\includegraphics[width = 0.2\linewidth]{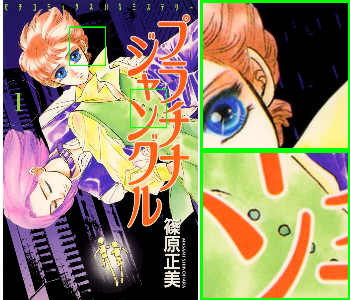}}
		\subfloat[LR~(23.25/0.8517)]{\includegraphics[width = 0.2\linewidth]{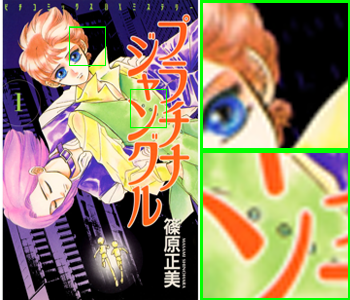}}
		\subfloat[LapSRN~(28.09/0.9456)]{\includegraphics[width = 0.2\linewidth]{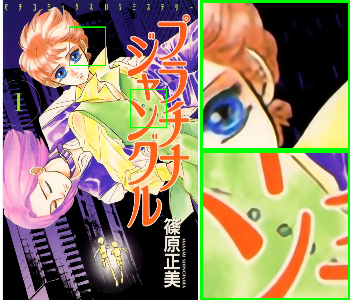}}
		\subfloat[IMDN~(29.64/0.9591)]{\includegraphics[width = 0.2\linewidth]{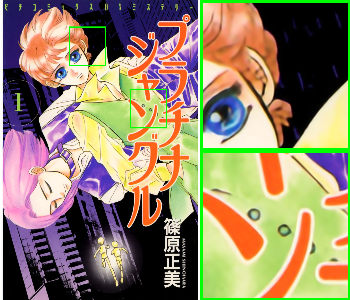}}
		\subfloat[Ours~(\textbf{29.86/0.9596})]{\includegraphics[width=0.2\linewidth]{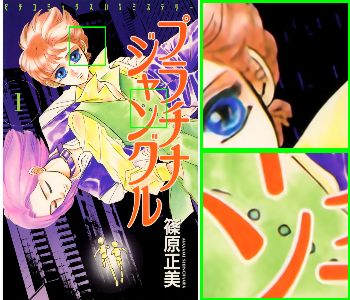}} \\
		
		\subfloat[HR~(PSNR/SSIM)]{\includegraphics[width = 0.2\linewidth]{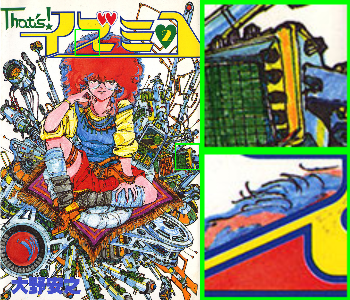}}
		\subfloat[LR~(16.78/0.5411)]{\includegraphics[width = 0.2\linewidth]{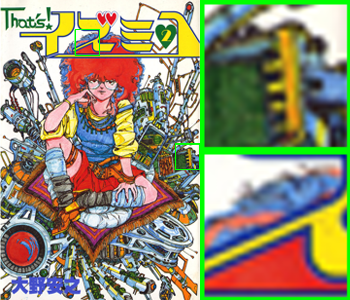}}
		\subfloat[LapSRN~(18.80/0.7019)]{\includegraphics[width = 0.2\linewidth]{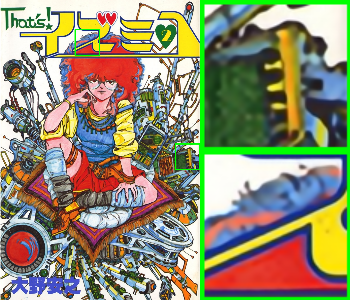}}
		\subfloat[IMDN~(19.64/0.7468)]{\includegraphics[width = 0.2\linewidth]{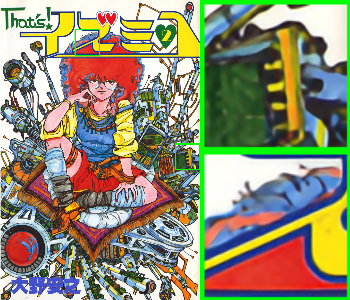}}
		\subfloat[Ours~(\textbf{19.70/0.7528})]{\includegraphics[width=0.2\linewidth]{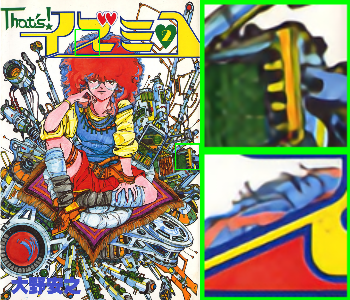}} \\
		
		\subfloat[HR~(PSNR/SSIM)]{\includegraphics[width = 0.2\linewidth]{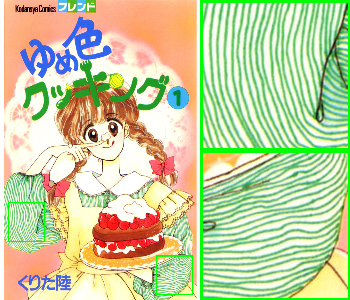}}
		\subfloat[LR~(22.90/0.7634)]{\includegraphics[width = 0.2\linewidth]{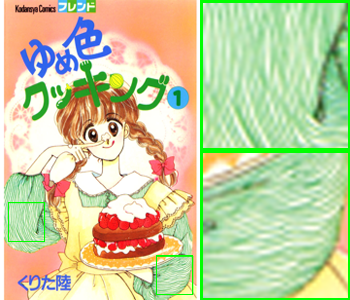}}
		\subfloat[LapSRN~(25.96/0.8850)]{\includegraphics[width = 0.2\linewidth]{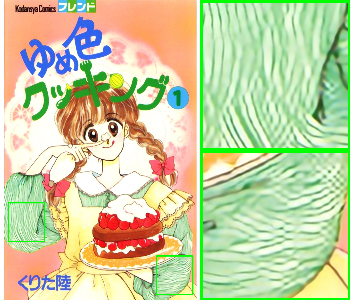}}
		\subfloat[IMDN~(26.56/0.9014)]{\includegraphics[width = 0.2\linewidth]{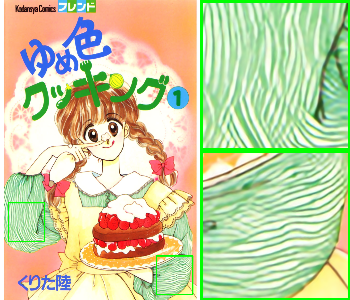}}
		\subfloat[Ours~(\textbf{27.12/0.9139})]{\includegraphics[width=0.2\linewidth]{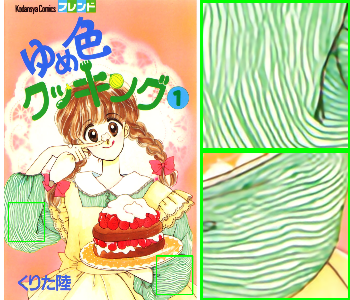}} \\
	\end{center}
	\vspace{-0.2cm}
	\caption{Visualization comparison on Manga109 dataset.}
	\label{fig:manga109}
\end{figure*}

According to Figure~\ref{fig:idg-detail}, we visualize the feature maps in four groups before fusing and the final feature map of the MFFG in Figure~\ref{fig:abl-distillation}. The input feature of MFFG is demonstrated in Figure~\ref{fig:abl-distillation} (a). Figures~\ref{fig:abl-distillation} (b)-(e) are the feature maps of groups from different CCBs, indicating features from $HC_0^3$ to $HC_3^0$ in Eq.~\ref{eq:idg}, respectively. Figure~\ref{fig:abl-distillation} (f) is the output of MFFG. All the images are produced by the same strategy. For each group, the feature maps are averaged and normalized in the range 0 to 1. We can obtain more and more refined edges and textures from (b) to (e), as the structural information is enhanced with the increase of CCBs. We take a zoomed-in green rectangle area as an example, which is guided by the red arrow. The small edge is clearer with enhanced contrast from (b) to (e). Thus, the experiment validates that the edges and structural textures are sharper in the output feature after the CCBs processing, which is obvious in the comparison between (a) and (f).

\subsubsection{Analysis on Network Structure}
We analysis and compare the performance of different network structure settings, including channel attention, and the different feature normalization mechanisms.

We demonstrate the effectiveness of the channel attention (CA) and feature normalization by testing the performance with or without two structures. The results are demonstrated in Table~\ref{tab:abl-ca}. We can see our model with CA and F-Norm achieves the best performance. The results show that model without F-Norm drops 0.06 dB on Manga109 benchmark, and the model without CA drops 0.05 dB on Urban100. In this point of view, CA and F-Norm are effective components for boosting the network performance.

\subsection{Comparison with State-of-the-Art Methods}
We compare Cross-SRN with three kinds of representative image SR works, including classical image SR works (SRCNN~\cite{srcnn_pami2016}, VDSR~\cite{vdsr_cvpr2016}, LapSRN~\cite{lapsrn_pami2019}, and CARN~\cite{carn_eccv2018}), multi-scale works (MRFN~\cite{mrfn_tmm2020} and IMDN~\cite{imdn_mm2019}), and structure-preserving works with edge map prior (DEGREE~\cite{degree_tip2017} and SeaNet-baseline~\cite{seanet_tip2020}). The PSNR/SSIM results are shown in Table~\ref{tab:BI-result}. All the methods for comparison follow the same training and testing protocol, and the performances are provided by the published papers. The last column of Table~\ref{tab:BI-result} demonstrates the proposed Cross-SRN achieves competitive or superior performances, and 90\% of the results achieve the bset (in bold) or second best (underline) performance. Especially, when compared with the classical and multi-scale works, Cross-SRN has near 0.1db PSNR improvement on Urban100 and Manga109, and achieves better performance on B100. The superior performance demonstrates that our network can recover structural textures more efficiently.

\begin{table*}[!t]
	\centering
	\caption{Average PSNR/SSIM with degradation model \textbf{BI} $\times2$, $\times3$, and $\times4$ on five benchmarks. The best and second performances are shown in \textbf{bold} and \underline{underline}.}
	\label{tab:BI-result}
	\begin{tabular}{|c|c|c|c|c|c|c|}
		\hline
		\multirow{2}{*}{Scale}& \multirow{2}{*}{Model}&  
		Set5~\cite{set5}& Set14~\cite{set14}& B100~\cite{b100}& Urban100~\cite{urban100}& Manga109~\cite{manga109} \\
		& & PSNR/SSIM & PSNR/SSIM & PSNR/SSIM & PSNR/SSIM & PSNR/SSIM\\
		\hline
		\multirow{18}{*}{$\times2$} &SRCNN~\cite{srcnn_pami2016}&
		36.66 / 0.9542& 32.42 / 0.9063& 31.36 / 0.8879& 29.50 / 0.8946& 35.74 / 0.9661\\
		
		& FSRCNN~\cite{fsrcnn_cvpr2016}&
		37.00 / 0.9558& 32.63 / 0.9088& 31.53 / 0.8920& 29.88 / 0.9020& 36.67 / 0.9694\\
		
		& VDSR~\cite{vdsr_cvpr2016}&
		37.53 / 0.9587& 33.03 / 0.9124& 31.90 / 0.8960& 30.76 / 0.9140& 37.22 / 0.9729\\
		
		& DRCN~\cite{drcn_cvpr2016}&
		37.63 / 0.9588& 33.04 / 0.9118& 31.85 / 0.8942& 30.75 / 0.9133& 37.63 / 0.9723\\
		
		& CNF~\cite{cnf_cvpr2017}&
		37.66 / 0.9590& 33.38 / 0.9136& 31.91 / 0.8962& - & - \\
		
		&LapSRN~\cite{lapsrn_cvpr2017}&
		37.52 / 0.9590& 33.08 / 0.9130& 31.80 / 0.8950& 30.41 / 0.9100& 37.27 / 0.9740\\
		
		&DRRN~\cite{drrn_cvpr2017}&
		37.74 / 0.9591& 33.23 / 0.9136& 32.05 / 0.8973& 31.23 / 0.9188& 37.92 / 0.9760\\
		
		&BTSRN~\cite{btsrn_cvpr2017}&
		37.75 / -& 33.20 / -& 32.05 / -& 31.63 / -& -\\
		
		&MemNet~\cite{memnet_iccv2017}&
		37.78 / 0.9597& 33.28 / 0.9142& 32.08 / 0.8978& 31.31 / 0.9195& 37.72 / 0.9740 \\
		
		&SelNet~\cite{selnet_cvpr2017}&
		37.89 / 0.9598& 33.61 / 0.9160& 32.08 / 0.8984& - &  - \\
		
		&CARN~\cite{carn_eccv2018}&
		37.76 / 0.9590& 33.52 / 0.9166& 32.09 / 0.8978& 31.92 / 0.9256& 38.36 / 0.9765\\
		
		&IMDN~\cite{imdn_mm2019}&
		\underline{38.00} / 0.9605& \textbf{33.63} / \underline{0.9177}& \underline{32.19 / 0.8996}& \underline{32.17 / 0.9283}& \textbf{38.88 / 0.9774}\\
		
		&RAN~\cite{ran_csvt2019}&
		37.58 / 0.9592 &33.10 / 0.9133 &31.92 / 0.8963& -& -\\
		
		&DNCL~\cite{dncl_csvt2019}&
		37.65 / 0.9599 &33.18 / 0.9141 &31.97 / 0.8971 &30.89 / 0.9158& - \\
		
		&FilterNet~\cite{filternet_csvt2020}&
		37.86 / 0.9610 &33.34 / 0.9150 &32.09 / 0.8990 &31.24 / 0.9200& - \\
		
		&MRFN~\cite{mrfn_tmm2020}&
		37.98 / 0.9611 &33.41 / 0.9159 &32.14 / 0.8997 &31.45 / 0.9221 &38.29 / 0.9759\\
		
		&SeaNet-baseline~\cite{seanet_tip2020}&
		37.99 / \underline{0.9607} & 33.60 / 0.9174 & 32.18 / 0.8995 & 32.08 / 0.9276 & 38.48 / 0.9768 \\
		
		&DEGREE~\cite{degree_tip2017}&
		37.58 / 0.9587& 33.06 / 0.9123& 31.80 / 0.8974& - & - \\

		&Cross-SRN~(Ours)&
		\textbf{38.03 / 0.9606}&\underline{33.62} / \textbf{0.9180}&\textbf{32.19 / 0.8997}&\textbf{32.28 / 0.9290}&\underline{38.75 / 0.9773} \\
		\hline

		\multirow{17}{*}{$\times3$}& SRCNN~\cite{srcnn_pami2016} &
		32.75 / 0.9090& 29.28 / 0.8209& 28.41 / 0.7863& 26.24 / 0.7989& 30.59 / 0.9107\\
		
		&FSRCNN~\cite{fsrcnn_cvpr2016}&
		33.16 / 0.9140& 29.43 / 0.8242& 28.53 / 0.7910& 26.43 / 0.8080& 30.98 / 0.9212\\
		
		&VDSR~\cite{vdsr_cvpr2016}&
		33.66 / 0.9213& 29.77 / 0.8314& 28.82 / 0.7976& 27.14 / 0.8279& 32.01 / 0.9310\\
		
		&DRCN~\cite{drcn_cvpr2016}&
		33.82 / 0.9226& 29.76 / 0.8311& 28.80 / 0.7963& 27.15 / 0.8276& 32.31 / 0.9328\\
		
		&CNF~\cite{cnf_cvpr2017}&
		33.74 / 0.9226& 29.90 / 0.8322& 28.82 / 0.7980& - & - \\
		
		&DRRN~\cite{drcn_cvpr2016}&
		34.03 / 0.9244& 29.96 / 0.8349& 28.95 / 0.8004& 27.53 / 0.8378& 32.74 / 0.9390\\
		
		&BTSRN~\cite{btsrn_cvpr2017}&
		34.03 / -& 29.90 / -& 28.97 / -& 27.75 / -& -\\
		
		&MemNet~\cite{memnet_iccv2017}&
		34.09 / 0.9248& 30.00 / 0.8350& 28.96 / 0.8001& 27.56 / 0.8376& 32.51 / 0.9369 \\
		
		&SelNet~\cite{selnet_cvpr2017}&
		34.27 / 0.9257& 30.30 / 0.8399& 28.97 / 0.8025& - &  - \\
		
		&CARN~\cite{carn_eccv2018}&
		34.29 / 0.9255& 30.29 / 0.8407& 29.06 / 0.8034& 28.06 / 0.8493& 33.50 / 0.9440\\
		
		&IMDN~\cite{imdn_mm2019}&
		34.36 / 0.9270& 30.32 / 0.8417& \underline{29.09 / 0.8046}& 28.17 / 0.8519& \underline{33.61 / 0.9445}\\
		
		&RAN~\cite{ran_csvt2019}&
		33.71 / 0.9223 & 29.84 / 0.8326 & 28.84 / 0.7981 & - & -\\
		
		&DNCL~\cite{dncl_csvt2019}&
		33.95 / 0.9232 & 29.93 / 0.8340 & 28.91 / 0.7995 & 27.27 / 0.8326 & -\\
		
		&FilterNet~\cite{filternet_csvt2020}&
		34.08 / 0.9250 & 30.03 / 0.8370 & 28.95 / 0.8030 & 27.55 / 0.8380 & -\\
		
		&MRFN~\cite{mrfn_tmm2020}&
		34.21 / 0.9267 & 30.03 / 0.8363 & 28.99 / 0.8029 & 27.53 / 0.8389 & 32.82 / 0.9396 \\
		
		&SeaNet-baseline~\cite{seanet_tip2020}&
		\underline{34.36 / 0.9280} & \textbf{30.34 / 0.8428} & \textbf{29.09 / 0.8053} & 28.17 / 0.8527 & 33.40 / 0.9444 \\
		
		&DEGREE~\cite{degree_tip2017}&
		33.76 / 0.9211& 29.82 / 0.8326& 28.74 / 0.7950& - & - \\
		
		&Cross-SRN~(Ours)&
		\textbf{34.43 / 0.9275}& \underline{30.33 / 0.8417}& \underline{29.09 / 0.8050}& \textbf{28.23 / 0.8535}& \textbf{33.65 / 0.9448}\\
		
		\hline
		\multirow{19}{*}{$\times4$}&SRCNN~\cite{srcnn_pami2016}&
		30.48 / 0.8628& 27.49 / 0.7503& 26.90 / 0.7101& 24.52 / 0.7221& 27.66 / 0.8505\\
		
		&FSRCNN~\cite{fsrcnn_cvpr2016}&
		30.71 / 0.8657& 27.59 / 0.7535& 26.98 / 0.7150& 24.62 / 0.7280& 27.90 / 0.8517\\
		
		&VDSR~\cite{vdsr_cvpr2016}&
		31.35 / 0.8838& 28.01 / 0.7674& 27.29 / 0.7251& 25.18 / 0.7524& 28.83 / 0.8809\\
		
		&DRCN~\cite{drcn_cvpr2016}&
		31.53 / 0.8854& 28.02 / 0.7670& 27.23 / 0.7233& 25.14 / 0.7510& 28.98 / 0.8816\\
		
		&CNF~\cite{cnf_cvpr2017}&
		31.55 / 0.8856& 28.15 / 0.7680& 27.32 / 0.7253& - & - \\
		
		&LapSRN~\cite{lapsrn_cvpr2017}&
		31.54 / 0.8850& 28.19 / 0.7720& 27.32 / 0.7280& 25.21 / 0.7560& 29.09 / 0.8845\\
		
		&DRRN~\cite{drcn_cvpr2016}&
		31.68 / 0.8888& 28.21 / 0.7720& 27.38 / 0.7284& 25.44 / 0.7638& 29.46 / 0.8960\\
		
		&BTSRN~\cite{btsrn_cvpr2017}&
		31.85 / -& 28.20 / -& 27.47 / -& 25.74 / -& -\\
		
		&MemNet~\cite{memnet_iccv2017}&
		31.74 / 0.8893& 28.26 / 0.7723& 27.40 / 0.7281& 25.50 / 0.7630& 29.42 / 0.8942 \\
		
		&SelNet~\cite{selnet_cvpr2017}&
		32.00 / 0.8931& 28.49 / 0.7783& 27.44 / 0.7325& - &  - \\
		
		&SRDenseNet~\cite{srdensenet_iccv2017}&
		32.02 / 0.8934& 28.50 / 0.7782& 27.53 / 0.7337& 26.05 / 0.7819& - \\
		
		&CARN~\cite{carn_eccv2018}&
		32.13 / 0.8937& \underline{28.60} / 0.7806& 27.58 / 0.7349& \underline{26.07} / 0.7837& \underline{30.47} / \underline{0.9084}\\
		
		&IMDN~\cite{imdn_mm2019}&
		\underline{32.21 / 0.8948}& 28.58 / \underline{0.7811}& 27.56 / 0.7353& 26.04 / \underline{0.7838}& {30.45} / 0.9075\\
		
		&RAN~\cite{ran_csvt2019}&
		31.43 / 0.8847 & 28.09 / 0.7691 & 27.31 / 0.7260 & - & -\\
		
		&DNCL~\cite{dncl_csvt2019}&
		31.66 / 0.8871 & 28.23 / 0.7717 & 27.39 / 0.7282 & 25.36 / 0.7606 & -\\
		
		&FilterNet~\cite{filternet_csvt2020}&
		31.74 / 0.8900 & 28.27 / 0.7730 & 27.39 / 0.7290 & 25.53 / 0.7680 & -\\
		
		&MRFN~\cite{mrfn_tmm2020}&
		31.90 / 0.8916 & 28.31 / 0.7746 & 27.43 / 0.7309 & 25.46 / 0.7654 & 29.57 / 0.8962 \\
		
		&SeaNet-baseline~\cite{seanet_tip2020}&
		32.18 / 0.8948 & \textbf{28.61 / 0.7822} & \underline{27.57 / 0.7359} & 26.05 / 0.7896 & 30.44 / \textbf{0.9088} \\
		
		&DEGREE~\cite{degree_tip2017}&
		31.47 / 0.8837& 28.10 / 0.7669& 27.20 / 0.7216& - & - \\

		&Cross-SRN~(Ours)&
		\textbf{32.24 / 0.8954}& \underline{28.59} / \underline{0.7817}& \textbf{27.58 / 0.7364}& \textbf{26.16 / 0.7881}& \textbf{30.53} / {0.9081}\\
		\hline
	\end{tabular}
\end{table*}

\begin{figure}[t]
	\begin{center}
		\subfloat[Param-PSNR comparison]{\includegraphics[width=0.48\linewidth]{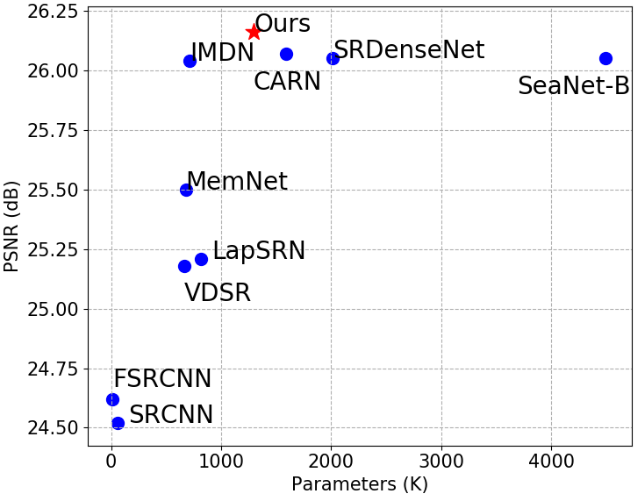}}
		\hspace{0.05cm}
		\subfloat[MACs-PSNR comparison]{\includegraphics[width=0.48\linewidth]{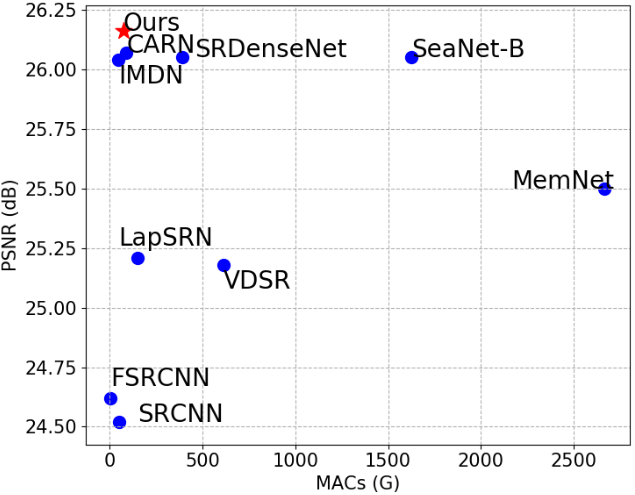}}
	\end{center}
	\caption{Comparisons of parameters, MACs and performances on Urban100 with \textbf{BI}$\times4$ degradation. (a) Parameter comparison. (b) MACs comparison.}
	\label{fig:mac-param}
\end{figure}

\begin{table*}[t]
	\centering
	\caption{Average PSNR/SSIM on selected benchmark with \textbf{BI}$\times4$ degradation.}
	\label{tab:selected}
	\begin{tabular}{|c|c|c|c|c|c|}
		\hline
		&LR& MS-LapSRN~\cite{lapsrn_pami2019}& VDSR~\cite{vdsr_cvpr2016}& IMDN~\cite{imdn_mm2019}& Cross-SRN \\
		\hline
		Urban100 (S)& 18.59 / 0.6356& 21.80 / 0.7916& 21.37 / 0.7696& 22.44 / 0.8074& 22.71 / 0.8145\\
		Manga109 (S)& 20.39 / 0.7314& 25.11 / 0.8795& -             & 25.96 / 0.8913& 26.03 / 0.8890\\
		Selected    & 19.54 / 0.6861& 23.55 / 0.8379& -             & 24.30 / 0.8515& 24.46 / 0.8537\\
		\hline
	\end{tabular}
\end{table*}

\begin{table*}[t]
	\centering
	\caption{Running time and PSNR/SSIM comparisons on Set5 with \textbf{BI}$\times4$ degradation.}
	\label{tab:time}
	\begin{tabular}{|c|c|c|c|c|c|}
		\hline
		\textbf{Method}& \textbf{LapSRN~\cite{lapsrn_pami2019}}	&\textbf{SRDenseNet~\cite{srdensenet_iccv2017}}	&\textbf{SeaNet (Baseline)~\cite{seanet_tip2020}}&	\textbf{CARN~\cite{carn_eccv2018}}	&\textbf{Ours}\\
		\hline
		\textbf{PSNR}& 				31.54&	32.02&	32.18&	32.13&	32.24\\
		\textbf{SSIM}& 				0.8850&	0.8934&	0.8948&	0.8937&	0.8954\\
		\textbf{Time Cost (ms)}& 	62.6&	304.8&	717.5&	41.0&	62.4\\
		\hline
	\end{tabular}
\end{table*}

The proposed Cross-SRN also has obvious advantage over edge map based methods DEGREE~\cite{degree_tip2017} and SeaNet-baseline~\cite{seanet_tip2020}. In Table \ref{tab:BI-result}, Cross-SRN achieves superior performances to DEGREE on all testing benchmarks. Compared with SeaNet, Cross-SRN achieves 90\% best or second best results over all the scale factors and benchmarks, while the percentage for SeaNet is only 43\%. On Urban100 and Manga109, the PSNR for Cross-SRN is 0.1-0.3 dB higher than SeaNet. It is worth noting that SeaNet-baseline contains 4.1M parameters, while Cross-SRN only contains around 1.3M. In this point of view, Cross-SRN can restore the structural information more effectively than other edge map based works with less parameters. 

We also compare our Cross-SRN with other works on the selective benchmark to show the effectiveness of structure information preservation. As shown in Table~\ref{tab:selected}, the selected 18 images from Urban100 are denoted as Urban100 (S), 20 images from Manga109 are denoted as Manga109 (S), and the total 38 images are denoted as Selected. The experiment result demonstrated Cross-SRN achieves superior performance to other works, which shows that our network has higher capacity on preserving structural information.

To demonstrate the restoration performance, visualization comparisons are shown in Figure~\ref{fig:set14}, Figure~\ref{fig:urban100} and Figure~\ref{fig:manga109}. The results on Set14 dataset demonstrate the effectiveness on structural information recovery. Urban100 benchmark is composed of high-resolution real world images with plentiful complex structural textures. Two representative instances of building are chosen to represent the restoration capacity. From Figure~\ref{fig:urban100}, Cross-SRN recover the grids and lines from tall buildings more efficiently. Compared with IMDN, Cross-SRN can prevent more texture mixtures, and regain more correct structural textures. It should be noted that IMDN is another image SR network with multi-scale fusion design. From this point of view, Cross-SRN has a convincing restoration capacity on structural information restoration.

Besides Urban100, we select two representative instances from Manga109 for comparison. Manga109 is a benchmark of comics with sharply defined areas and much high-frequency information. We compare Cross-SRN with LR, LapSRN and IMDN. The results are shown in Figure~\ref{fig:manga109}. From the comparisons, Cross-SRN can restore the high-frequency information more accurately. Specially, the tiny textures can be recovered more effectively by Cross-SRN, such as areas of the word and the eye. Meanwhile, the serried line textures, which are mixed by down-sampling, can be well recovered by Cross-SRN. From the visualization comparison, Cross-SRN gains superior restoration performance than other lightweight works.

It is worth noting that the proposed Cross-SRN achieves superior performance with fewer parameters and lower computation complexity than other methods. Figure~\ref{fig:mac-param} demonstrates an intuitive comparison on the PSNR and corresponding parameters and MACs on Urban100 dataset. Our method is labeled by red star, while other methods are labeled by blue points. The proposed method achieves the best performance with the red star over all the blue points with less paramethers and lower MACs. Thus, Cross-SRN proves to be an efficient design for structure-preserving image restoration.

Besides the objective and subjective comparisons, we also compare the speed of different methods. We calculate the time cost by restoring a random 720P image with $\textbf{BI}\times4$ degradation for 100 times, and choose the average value for comparison. The results are shown in Table~\ref{tab:time}. In the table, our network achieves the best PSNR/SSIM performance with a fast speed. Compared with CARN, we achieve 0.1 dB PSNR improvement and only require 20 ms more time cost.

\section{Conclusion}
In this paper, we propose a network termed as Cross-SRN for edge-preserving image super-resolution. Inspired by edge detection methods, a novel cross convolution operation is designed to exploit the structural information more effectively. The cross convolution leverages two perpendicular factorized filters parallelly to increase the matrix rank and preserve more information. Based on the cross convolution, CCBs are investigated with CA and F-Norm to concern the inherent correlation of channel-wise and spatial features separately. MFFG groups CCBs in a multi-scale feature fusion manner for efficient hierarchical feature exploration. Experimental result shows that Cross-SRN has demonstrated superior restoration capacity than other lightweight works on both quantitative and qualitative comparisons.

\section*{Acknowledgement}
This work is partially supported by the NSF of China under grand Nos. 62088102, 62031013 and High-performance Computing Platform of Peking University, which are gratefully acknowledged. This work is also partially supported by the NSF of China under grant Nos. 61876030, 61733002.

\bibliographystyle{ieeetr}
\bibliography{main, edgesr, review}

\begin{thebibliography}{10}

\bibitem{gao2021digital}
W.~Gao, S.~Ma, L.~Duan, Y.~Tian, P.~Xing, Y.~Wang, S.~Wang, H.~Jia, and
  T.~Huang, ``Digital retina: A way to make the city brain more efficient by
  visual coding,'' {\em IEEE Transactions on Circuits and Systems for Video
  Technology}, vol.~31, no.~11, pp.~4147--4161, 2021.

\bibitem{srcnn_pami2016}
C.~{Dong}, C.~C. {Loy}, K.~{He}, and X.~{Tang}, ``Image super-resolution using
  deep convolutional networks,'' {\em IEEE Transactions on Pattern Analysis and
  Machine Intelligence}, vol.~38, no.~2, pp.~295--307, 2016.

\bibitem{drn_cvpr2020}
Y.~Guo, J.~Chen, J.~Wang, Q.~Chen, J.~Cao, Z.~Deng, Y.~Xu, and M.~Tan,
  ``Closed-loop matters: Dual regression networks for single image
  super-resolution,'' in {\em {IEEE/CVF} Conference on Computer Vision and
  Pattern Recognition}, pp.~5406--5415, 2020.

\bibitem{han_eccv2020}
B.~Niu, W.~Wen, W.~Ren, X.~Zhang, L.~Yang, S.~Wang, K.~Zhang, X.~Cao, and
  H.~Shen, ``Single image super-resolution via a holistic attention network,''
  in {\em European Conference on Computer Vision}, vol.~12357, pp.~191--207,
  2020.

\bibitem{latticenet_eccv2020}
X.~Luo, Y.~Xie, Y.~Zhang, Y.~Qu, C.~Li, and Y.~Fu, ``Latticenet: Towards
  lightweight image super-resolution with lattice block,'' in {\em European
  Conference on Computer Vision}, Springer, 2020.

\bibitem{sobel}
Y.~Chien, ``Pattern classification and scene analysis,'' {\em IEEE Transactions
  on Automatic Control}, vol.~19, no.~4, pp.~462--463, 1974.

\bibitem{canny_pani1986}
J.~{Canny}, ``A computational approach to edge detection,'' {\em IEEE
  Transactions on Pattern Analysis and Machine Intelligence}, vol.~PAMI-8,
  no.~6, pp.~679--698, 1986.

\bibitem{bdcn_pami2020}
J.~{He}, S.~{Zhang}, M.~{Yang}, Y.~{Shan}, and T.~{Huang}, ``Bdcn:
  Bi-directional cascade network for perceptual edge detection,'' {\em IEEE
  Transactions on Pattern Analysis and Machine Intelligence}, pp.~1--1, 2020.

\bibitem{rcf_pami2019}
Y.~{Liu}, M.~{Cheng}, X.~{Hu}, J.~{Bian}, L.~{Zhang}, X.~{Bai}, and J.~{Tang},
  ``Richer convolutional features for edge detection,'' {\em IEEE Transactions
  on Pattern Analysis and Machine Intelligence}, vol.~41, no.~8,
  pp.~1939--1946, 2019.

\bibitem{r1q1_1}
X.~Ran and N.~Farvardin, ``A perceptually motivated three-component image
  model-part i: description of the model,'' {\em IEEE Transactions on Image
  Processing}, vol.~4, no.~4, pp.~401--415, 1995.

\bibitem{r1q1_2}
H.~Wang, X.~Hu, X.~Zhao, and Y.~Zhang, ``Wide weighted attention multi-scale
  network for accurate mr image super-resolution,'' {\em IEEE Transactions on
  Circuits and Systems for Video Technology}, pp.~1--1, 2021.

\bibitem{r1q1_3}
S.~Mandal and A.~K. Sao, ``Edge preserving single image super resolution in
  sparse environment,'' in {\em 2013 IEEE International Conference on Image
  Processing}, pp.~967--971, 2013.

\bibitem{r1q1_4}
Y.~Li, J.~Liu, W.~Yang, and Z.~Guo, ``Neighborhood regression for
  edge-preserving image super-resolution,'' in {\em 2015 IEEE International
  Conference on Acoustics, Speech and Signal Processing (ICASSP)},
  pp.~1201--1205, 2015.

\bibitem{r1q1_5}
S.~Huang, J.~Sun, Y.~Yang, Y.~Fang, P.~Lin, and Y.~Que, ``Robust single-image
  super-resolution based on adaptive edge-preserving smoothing
  regularization,'' {\em IEEE Transactions on Image Processing}, vol.~27,
  no.~6, pp.~2650--2663, 2018.

\bibitem{r1q1_6}
S.~Pelletier and J.~R. Cooperstock, ``Preconditioning for edge-preserving image
  super resolution,'' {\em IEEE Transactions on Image Processing}, vol.~21,
  no.~1, pp.~67--79, 2012.

\bibitem{r1q1_7}
L.~Wang, S.~Xiang, G.~Meng, H.~Wu, and C.~Pan, ``Edge-directed single-image
  super-resolution via adaptive gradient magnitude self-interpolation,'' {\em
  IEEE Transactions on Circuits and Systems for Video Technology}, vol.~23,
  no.~8, pp.~1289--1299, 2013.

\bibitem{seanet_tip2020}
F.~{Fang}, J.~{Li}, and T.~{Zeng}, ``Soft-edge assisted network for single
  image super-resolution,'' {\em IEEE Transactions on Image Processing},
  vol.~29, pp.~4656--4668, 2020.

\bibitem{degree_tip2017}
W.~{Yang}, J.~{Feng}, J.~{Yang}, F.~{Zhao}, J.~{Liu}, Z.~{Guo}, and S.~{Yan},
  ``Deep edge guided recurrent residual learning for image super-resolution,''
  {\em IEEE Transactions on Image Processing}, vol.~26, no.~12, pp.~5895--5907,
  2017.

\bibitem{urban100}
J.~{Huang}, A.~{Singh}, and N.~{Ahuja}, ``Single image super-resolution from
  transformed self-exemplars,'' in {\em IEEE Conference on Computer Vision and
  Pattern Recognition}, pp.~5197--5206, 2015.

\bibitem{lapsrn_cvpr2017}
W.~{Lai}, J.~{Huang}, N.~{Ahuja}, and M.~{Yang}, ``Deep laplacian pyramid
  networks for fast and accurate super-resolution,'' in {\em IEEE Conference on
  Computer Vision and Pattern Recognition}, pp.~5835--5843, 2017.

\bibitem{carn_eccv2018}
N.~Ahn, B.~Kang, and K.~Sohn, ``Fast, accurate, and lightweight
  super-resolution with cascading residual network,'' in {\em European
  Conference on Computer Vision}, vol.~11214, pp.~256--272, 2018.

\bibitem{msrn_eccv2018}
J.~Li, F.~Fang, K.~Mei, and G.~Zhang, ``Multi-scale residual network for image
  super-resolution,'' in {\em European Conference on Computer Vision},
  vol.~11212, pp.~527--542, 2018.

\bibitem{isrn}
Y.~Liu, S.~Wang, J.~Zhang, S.~Wang, S.~Ma, and W.~Gao, ``Iterative network for
  image super-resolution,'' {\em IEEE Transactions on Multimedia}, pp.~1--1,
  2021.

\bibitem{senet_pami2020}
J.~{Hu}, L.~{Shen}, S.~{Albanie}, G.~{Sun}, and E.~{Wu},
  ``Squeeze-and-excitation networks,'' {\em IEEE Transactions on Pattern
  Analysis and Machine Intelligence}, vol.~42, no.~8, pp.~2011--2023, 2020.

\bibitem{vdsr_cvpr2016}
J.~{Kim}, J.~K. {Lee}, and K.~M. {Lee}, ``Accurate image super-resolution using
  very deep convolutional networks,'' in {\em IEEE Conference on Computer
  Vision and Pattern Recognition}, pp.~1646--1654, 2016.

\bibitem{edsr_cvpr2017}
B.~{Lim}, S.~{Son}, H.~{Kim}, S.~{Nah}, and K.~M. {Lee}, ``Enhanced deep
  residual networks for single image super-resolution,'' in {\em IEEE
  Conference on Computer Vision and Pattern Recognition Workshops},
  pp.~1132--1140, 2017.

\bibitem{lapsrn_pami2019}
W.~{Lai}, J.~{Huang}, N.~{Ahuja}, and M.~{Yang}, ``Fast and accurate image
  super-resolution with deep laplacian pyramid networks,'' {\em IEEE
  Transactions on Pattern Analysis and Machine Intelligence}, vol.~41, no.~11,
  pp.~2599--2613, 2019.

\bibitem{rcan_eccv2018}
Y.~Zhang, K.~Li, K.~Li, L.~Wang, B.~Zhong, and Y.~Fu, ``Image super-resolution
  using very deep residual channel attention networks,'' in {\em European
  Conference on Computer Vision}, vol.~11211, pp.~294--310, 2018.

\bibitem{rdn_pami2020}
Y.~{Zhang}, Y.~{Tian}, Y.~{Kong}, B.~{Zhong}, and Y.~{Fu}, ``Residual dense
  network for image restoration,'' {\em IEEE Transactions on Pattern Analysis
  and Machine Intelligence}, pp.~1--1, 2020.

\bibitem{spsr_cvpr2020}
C.~Ma, Y.~Rao, Y.~Cheng, C.~Chen, J.~Lu, and J.~Zhou, ``Structure-preserving
  super resolution with gradient guidance,'' in {\em {IEEE/CVF} Conference on
  Computer Vision and Pattern Recognition}, pp.~7766--7775, 2020.

\bibitem{idn_cvpr2018}
Z.~{Hui}, X.~{Wang}, and X.~{Gao}, ``Fast and accurate single image
  super-resolution via information distillation network,'' in {\em IEEE/CVF
  Conference on Computer Vision and Pattern Recognition}, pp.~723--731, 2018.

\bibitem{rfdn_eccv2020}
J.~Liu, J.~Tang, and G.~Wu, ``Residual feature distillation network for
  lightweight image super-resolution,'' in {\em European Conference on Computer
  Vision}, vol.~12537, pp.~41--55, 2020.

\bibitem{san_cvpr2019}
T.~{Dai}, J.~{Cai}, Y.~{Zhang}, S.~{Xia}, and L.~{Zhang}, ``Second-order
  attention network for single image super-resolution,'' in {\em IEEE/CVF
  Conference on Computer Vision and Pattern Recognition}, pp.~11057--11066,
  2019.

\bibitem{imdn_mm2019}
Z.~Hui, X.~Gao, Y.~Yang, and X.~Wang, ``Lightweight image super-resolution with
  information multi-distillation network,'' in {\em ACM International
  Conference on Multimedia}, p.~2024–2032, 2019.

\bibitem{inceptionV3}
C.~Szegedy, V.~Vanhoucke, S.~Ioffe, J.~Shlens, and Z.~Wojna, ``Rethinking the
  inception architecture for computer vision,'' in {\em IEEE Conference on
  Computer Vision and Pattern Recognition (CVPR)}, pp.~2818--2826, 2016.

\bibitem{separable}
A.~Gholami, K.~Kwon, B.~Wu, Z.~Tai, X.~Yue, P.~Jin, S.~Zhao, and K.~Keutzer,
  ``Squeezenext: Hardware-aware neural network design,'' in {\em IEEE/CVF
  Conference on Computer Vision and Pattern Recognition Workshops (CVPRW)},
  pp.~1719--171909, 2018.

\bibitem{div2k}
E.~{Agustsson} and R.~{Timofte}, ``Ntire 2017 challenge on single image
  super-resolution: Dataset and study,'' in {\em IEEE Conference on Computer
  Vision and Pattern Recognition Workshops}, pp.~1122--1131, 2017.

\bibitem{bicubic}
R.~{Keys}, ``Cubic convolution interpolation for digital image processing,''
  {\em IEEE Transactions on Acoustics, Speech, and Signal Processing}, vol.~29,
  no.~6, pp.~1153--1160, 1981.

\bibitem{adam_iclr2015}
D.~P. Kingma and J.~Ba, ``Adam: {A} method for stochastic optimization,'' in
  {\em International Conference on Learning Representations}, 2015.

\bibitem{set5}
M.~Bevilacqua, A.~Roumy, C.~Guillemot, and M.~line Alberi~Morel,
  ``Low-complexity single-image super-resolution based on nonnegative neighbor
  embedding,'' in {\em British Machine Vision Conference}, pp.~135.1--135.10,
  2012.

\bibitem{set14}
R.~Zeyde, M.~Elad, and M.~Protter, ``On single image scale-up using
  sparse-representations,'' in {\em Curves and Surfaces}, (Berlin, Heidelberg),
  pp.~711--730, Springer Berlin Heidelberg, 2012.

\bibitem{b100}
D.~{Martin}, C.~{Fowlkes}, D.~{Tal}, and J.~{Malik}, ``A database of human
  segmented natural images and its application to evaluating segmentation
  algorithms and measuring ecological statistics,'' in {\em IEEE International
  Conference on Computer Vision}, vol.~2, pp.~416--423 vol.2, 2001.

\bibitem{manga109}
Y.~Matsui, K.~Ito, Y.~Aramaki, A.~Fujimoto, T.~Ogawa, T.~Yamasaki, and
  K.~Aizawa, ``Sketch-based manga retrieval using manga109 dataset,'' {\em
  Multimedia Tools and Applications}, vol.~76, no.~20, pp.~21811--21838, 2017.

\bibitem{mrfn_tmm2020}
Z.~He, Y.~Cao, L.~Du, B.~Xu, J.~Yang, Y.~Cao, S.~Tang, and Y.~Zhuang, ``{MRFN:}
  multi-receptive-field network for fast and accurate single image
  super-resolution,'' {\em {IEEE} Transactions on Multimedia}, vol.~22, no.~4,
  pp.~1042--1054, 2020.

\bibitem{fsrcnn_cvpr2016}
C.~Dong, C.~C. Loy, and X.~Tang, ``Accelerating the super-resolution
  convolutional neural network,'' in {\em European Conference on Computer
  Vision}, pp.~391--407, 2016.

\bibitem{drcn_cvpr2016}
J.~Kim, J.~K. Lee, and K.~M. Lee, ``Deeply-recursive convolutional network for
  image super-resolution,'' in {\em {IEEE} Conference on Computer Vision and
  Pattern Recognition}, pp.~1637--1645, 2016.

\bibitem{cnf_cvpr2017}
H.~Ren, M.~El{-}Khamy, and J.~Lee, ``Image super resolution based on fusing
  multiple convolution neural networks,'' in {\em {IEEE} Conference on Computer
  Vision and Pattern Recognition Workshops}, pp.~1050--1057, 2017.

\bibitem{drrn_cvpr2017}
Y.~Tai, J.~Yang, and X.~Liu, ``Image super-resolution via deep recursive
  residual network,'' in {\em {IEEE} Conference on Computer Vision and Pattern
  Recognition}, pp.~2790--2798, 2017.

\bibitem{btsrn_cvpr2017}
Y.~Fan, H.~Shi, J.~Yu, D.~Liu, W.~Han, H.~Yu, Z.~Wang, X.~Wang, and T.~S.
  Huang, ``Balanced two-stage residual networks for image super-resolution,''
  in {\em {IEEE} Conference on Computer Vision and Pattern Recognition
  Workshops}, pp.~1157--1164, 2017.

\bibitem{memnet_iccv2017}
Y.~Tai, J.~Yang, X.~Liu, and C.~Xu, ``Memnet: {A} persistent memory network for
  image restoration,'' in {\em {IEEE} International Conference on Computer
  Vision}, pp.~4549--4557, 2017.

\bibitem{selnet_cvpr2017}
J.~{Choi} and M.~{Kim}, ``A deep convolutional neural network with selection
  units for super-resolution,'' in {\em IEEE Conference on Computer Vision and
  Pattern Recognition Workshops}, pp.~1150--1156, 2017.

\bibitem{ran_csvt2019}
Y.~Wang, L.~Wang, H.~Wang, and P.~Li, ``Resolution-aware network for image
  super-resolution,'' {\em {IEEE} Transactions on Circuits and Systems for
  Video Technology}, vol.~29, no.~5, pp.~1259--1269, 2019.

\bibitem{dncl_csvt2019}
C.~Xie, W.~Zeng, and X.~Lu, ``Fast single-image super-resolution via deep
  network with component learning,'' {\em {IEEE} Transactions on Circuits and
  Systems for Video Technology}, vol.~29, no.~12, pp.~3473--3486, 2019.

\bibitem{filternet_csvt2020}
F.~Li, H.~Bai, and Y.~Zhao, ``Filternet: Adaptive information filtering network
  for accurate and fast image super-resolution,'' {\em {IEEE} Transactions on
  Circuits and Systems for Video Technology}, vol.~30, no.~6, pp.~1511--1523,
  2020.

\bibitem{srdensenet_iccv2017}
T.~Tong, G.~Li, X.~Liu, and Q.~Gao, ``Image super-resolution using dense skip
  connections,'' in {\em {IEEE} International Conference on Computer Vision},
  pp.~4809--4817, 2017.

\end{thebibliography}

\end{document}